\documentclass[aps,prc,a4paper,nofootinbib,
preprintnumbers,superscriptaddress,showpacs,showkeys,twocolumn]{revtex4}

\usepackage{amsmath}
\usepackage{amssymb}
\usepackage{bm}
\usepackage{graphicx}
\usepackage{color}

\newcommand{\be}{\begin{equation}}
\newcommand{\ee}{\end{equation}}
\newcommand{\ba}{\begin{eqnarray}}
\newcommand{\ea}{\end{eqnarray}}

\begin{document}

\title{Diffusion of heavy quarks in the early stage of high energy nuclear collisions
at RHIC and LHC}

\author{J. H. Liu}
\affiliation{School of Nuclear Science and Technology, Lanzhou University, 222 South Tianshui Road, Lanzhou 730000, China}
\affiliation{INFN-Laboratori Nazionali del Sud, Via S. Sofia 62, I-95123 Catania, Italy}

\author{S. Plumari}
\affiliation{Department of Physics and Astronomy, University of Catania, Via S. Sofia 64, I-95125 Catania}
\affiliation{INFN-Laboratori Nazionali del Sud, Via S. Sofia 62, I-95123 Catania, Italy}

\author{S. K. Das}
\affiliation{School of Physical Sciences, Indian Institute of Technology Goa, Ponda-403401, Goa, India}

\author{V. Greco}
\affiliation{Department of Physics and Astronomy, University of Catania, Via S. Sofia 64, I-95125 Catania}
\affiliation{INFN-Laboratori Nazionali del Sud, Via S. Sofia 62, I-95123 Catania, Italy}

\author{M. Ruggieri}\email{ruggieri@lzu.edu.cn}
\affiliation{School of Nuclear Science and Technology, Lanzhou University, 222 South Tianshui Road, Lanzhou 730000, China}


\begin{abstract}
We study the diffusion of charm and beauty in the early stage of high energy nuclear collisions
at RHIC and LHC energies, considering the
interaction of these heavy quarks with the evolving Glasma by means of the Wong equations,
in the color-$SU(2)$ case.
In comparison with previous works, we add the longitudinal expansion of the gluon medium as well as we estimate
the diffusion coefficient in momentum space, and the effect of energy loss due to gluon radiation.
We find that heavy quarks diffuse in the strong transverse color fields in the very early stage (0.2-0.3 fm/c)
and this leads to a suppression at low $p_T$ and enhancement at intermediate low $p_T$.
The shape of the observed nuclear suppression factor obtained within our calculations is in qualitative
agreement with the experimental results of the same quantity for $D-$mesons in proton-nucleus collisions.
We compute the nuclear suppression factor in nucleus-nucleus collisions as well,
for both charm and beauty, finding a substantial impact of the evolving Glasma phase on these, suggesting
that initialization of heavy quarks spectra in the quark-gluon plasma phase should not neglect
the early evolution in the strong gluon fields.
\end{abstract}



\maketitle

\section{Introduction}

The study of the initial condition in high energy collisions
is an interesting problem related to the physics of
relativistic heavy ion collisions (RHICs), as well as to that of high energy proton-nucleus (pA) and
proton-proton (pp) collisions. At very high collision energy the two colliding nuclei
can be described within the color-glass-condensate (CGC) effective theory
\cite{McLerran:1993ni,McLerran:1993ka,McLerran:1994vd,Gelis:2010nm,Iancu:2003xm,McLerran:2008es,Gelis:2012ri},
in which fast partons dynamic is frozen by time dilatation and they act as static sources
for low momentum gluons: their large occupation number allows to treat them as classical fields.
The collision of the two CGC sheets, each representing one of the
colliding objects in high energy collisions, leads to the formation of strong gluon fields in the forward light cone
called the Glasma
\cite{Kovner:1995ja,Kovner:1995ts,Gyulassy:1997vt,Lappi:2006fp,Fukushima:2006ax,
Fries:2006pv,Chen:2015wia,Fujii:2008km,Krasnitz:2000gz,Krasnitz:2003jw,Krasnitz:2001qu}.
The Glasma consists of longitudinal color-electric and color-magnetic fields in the weak 
coupling regime and characterized by large gluon occupation number,
$A_\mu^a \simeq 1/g$ with $g$ the QCD coupling,
so they are described by classical field theory namely the Classical Yang-Mills (CYM) theory.
Finite coupling bring up fluctuations on the top of the Glasma \cite{Romatschke:2005pm,Romatschke:2006nk,Fukushima:2011nq,
Fukushima:2013dma,Iida:2014wea,Gelis:2013rba,Epelbaum:2013waa,Tanji:2011di,Ryblewski:2013eja,Ruggieri:2015yea,
Berges:2012cj,Berges:2013fga,Berges:2013lsa,Berges:2013eia,Ruggieri:2017ioa} that we
do not consider in the present manuscript.

Heavy quarks are considered as nobel probes of the system produced in high energy nuclear collisions,
both for the pre-equilibrium phase and for the thermalized quark-gluon plasma (QGP), see
\cite{Rapp:2018qla, Aarts:2016hap,Dong:2019unq,Cao:2018ews, Das:2015ana, Das:2013kea,Das:2016cwd,
Das:2017dsh,Das:2015aga,Beraudo:2015wsd,Xu:2015iha,Ozvenchuk:2017ojj,
Prino:2016cni,Andronic:2015wma,Chandra:2015gma,Mrowczynski:2017kso,
Ruggieri:2018rzi, Ruggieri:2018ies,Sun:2019fud}  and references therein.
Their formation time is very small in comparison with the one of light quarks:
indeed, this can be estimated as  $\tau_\mathrm{form} \approx 1/(2m)$ with
$m$ the heavy quark mass which gives $\tau_\mathrm{form} \leq 0.1$ fm/c for charm and beauty quarks.
Since heavy quarks are produced
immediately after the collision, they can propagate in the evolving Glasma fields and probe its evolution.

Diffusion of heavy quarks in the early stage of high energy nuclear collisions 
has been studied previously in \cite{Mrowczynski:2017kso}
within a simplified approach based on a Fokker-Planck equation. 
A similar study has been performed in~\cite{Ruggieri:2018rzi,Ruggieri:2018ies}
where emphasis has been put on the calculation of the nuclear modification factor in p-Pb collision, $R_\mathrm{pPb}$. 
In comparison with \cite{Mrowczynski:2017kso}, the work in  \cite{Ruggieri:2018rzi,Ruggieri:2018ies}
goes beyond the small momentum exchange approximation,
solving  the heavy quarks dynamics in the evolving Glasma fields via the Wong equations;
this method is equivalent to solve the Boltzmann-Vlasov equations
for the heavy quarks in a collisionless plasma. As a matter of fact,
the Boltzmann-Vlasov equations can be solved by means of the test particles method
which analogous to solve the classical equations of motion of the test particles, here represented by the heavy quarks,
and these classical equations are the Wong equations.
Then in \cite{Sun:2019fud}  the effect of this dynamics on the elliptic flow in Pb-Pb collisions has been considered.
However, in \cite{Ruggieri:2018rzi,Ruggieri:2018ies,Sun:2019fud}  the longitudinal expansion of the medium
has been ignored. 

The main purpose of the present article is to report on
a more complete study of the problem of diffusion of charm and beauty in the early stage
of high energy nuclear collisions, improving the work already presented in  \cite{Ruggieri:2018rzi,Sun:2019fud}
by including the longitudinal expansion of the gluon medium.
For the sake of computational simplicity we consider the color-$SU(2)$ case here.
Although the expansion dilutes the energy density and makes the effect on the modification factors
smaller than the one found in \cite{Ruggieri:2018rzi,Sun:2019fud}, we still find a substantial contribution 
of the early stage to these quantities both in pA and in AA collisions.

In addition to this necessary improvement, we also estimate 
the diffusion coefficient in momentum space of the heavy quarks in the evolving Glasma,
which was not done in  \cite{Ruggieri:2018rzi,Sun:2019fud};
finally, we present an estimate
of the energy loss due to gluon radiation that acts as a drag force on the heavy quarks.
We find that unless we artificially take a low average energy density in the full evolution,
the effect of the drag force is not strong enough to cancel the tilt of the spectrum of charm and beauty.
We remark that we do not include cold nuclear matter effects in our calculations
\cite{Eskola:2009uj,Fujii:2013yja,Ducloue:2015gfa,Rezaeian:2012ye,
Albacete:2013ei,Albacete:2016veq,Prino:2016cni,Andronic:2015wma}:
inclusion of these will be important, in particular, in pA collisions in the forward rapidity region;
we will consider these in future studies.

The plan of the paper is as follows: in Section II we review the Glasma and the Classical Yang-Mills equations;
in Section III we review the Wong equations; in Section IV we discuss our results for $R_\mathrm{pPb}$ and
$R_\mathrm{PbPb}$; in Section V we estimate the effect of energy loss due to gluon radiation; finally,
in Section VI we draw our conclusions.

\section{Glasma and classical Yang-Mills equations}
In this section, we briefly describe the McLerran-Venugopalan (MV)
model \cite{McLerran:1993ni,McLerran:1993ka,McLerran:1994vd,Kovchegov:1996ty} which is used to prepare
the initial condition known as the Glasma.
We mention that in our notation the gauge fields have been rescaled by the QCD coupling
$A_\mu \rightarrow A_\mu/g$.

In the MV model, the static color charge densities $\rho_a$
on the nucleus $A$, the colliding objects, are assumed to be random variables that are
normally distributed with zero mean
and variance described by the equation
\begin{equation}
\langle \rho^a_A(\bm x_T)\rho^b_A(\bm y_T)\rangle =
(g^2\mu_A)^2 \varphi_A(\bm x_T)\delta^{ab}\delta^{(2)}(\bm x_T-\bm y_T);
\label{eq:dfg}
\end{equation}
here $A$ corresponds to either the Pb nucleus or the  proton, $a$ and $b$ denote the adjoint color index.
In this work we limit ourselves, for simplicity, to the case of the $SU(2)$ color group therefore
$a,b=1,2,3$.
In Eq.~(\ref{eq:dfg}) $g^2\mu_A$ indicates the color charge density and it is of the order 
of the saturation momentum $Q_s$ \cite{Lappi:2007ku}.

The function $\varphi_A(\bm x_T)$  in Eq.~\eqref{eq:dfg} allows for a nonuniform probability distribution
of the color charges in the transverse plane.
For the Pb nucleus we assume a uniform
probability and take $\varphi(\bm x_T)=1$. For the proton we use a gaussian  $\varphi_A(\bm x_T)$
that mimics the distribution of color charges obtained after rapidity evolution from the  
constituent quark model \cite{Schenke:2014zha,Schenke:2015aqa,Mantysaari:2017cni,Mantysaari:2016jaz}, namely
\begin{equation}
\varphi_p(\bm x_T) = e^{-\bm x_T^2/2 B}.\label{eq:xi_distri}
\end{equation}
The parameter in Eq.~\eqref{eq:xi_distri} is
$B = 3$ GeV$^{-2}$.   
  
For the proton $g^2\mu_p \varphi_p(\bm x_T)^{1/2}$
can be understood as an $\bm x_T-$dependent $g^2\mu$.
Following the result of \cite{Lappi:2007ku},  we fix $g^2\mu_p$
for each event assuming that $\langle  g^2\mu_p \varphi_p(\bm x_T)^{1/2}\rangle/ Q_s= 0.57$
where the average is defined with $\varphi_p(\bm x_T)$ as a weight function,
then estimating
$Q_s$ at the relevant energy by using
the standard GBW fit  \cite{GolecBiernat:1999qd,GolecBiernat:1998js,Kovchegov:2012mbw}
\begin{equation}
Q_s^2 = Q_{s,0}^2 \left(\frac{x_0}{x}\right)^\lambda,\label{eq:QS_slide}
\end{equation}
where $\lambda=0.277$, $Q_0=1$ GeV and $x_0=4.1\times 10^{-5}$.
The relevant value of $x$ for the two colliding objects can be obtained at midrapidity as $\langle p_T\rangle/\sqrt{s}$
where $\langle p_T\rangle$ corresponds to the average $p_T$ of the gluons produced
in that  collision.
For example, at the RHIC colliding energy for $x=0.01$
we find $Q_s = 0.47$ GeV in agreement with the estimate of \cite{Albacete:2012xq}.
At the LHC energy, $\sqrt{s}=5.02$ TeV, we obtain $Q_s = 0.80$ GeV which gives
$\langle  g^2\mu_p \varphi_p(\bm x_T)^{1/2}\rangle = 1.41$ GeV.

For the Pb nucleus the uncertainty on the $Q_s$ and on $g^2\mu$ comes from
the different model used to estimate $Q_s$ for a large nucleus. In this case the GBW fit is modified as
\begin{equation}
Q_s^2 = f(A)Q_{s,0}^2 \left(\frac{x_0}{x}\right)^\lambda,\label{eq:QS_slide2}
\end{equation}
where
\begin{equation}
f(A) = A^{1/3}
\end{equation}
within a naive scaling hypothesis , and
\begin{equation}
f(A) =c A^{1/3}\log A
\end{equation}
within the IP-Sat model \cite{Kowalski:2007rw}.
While other forms of $f(A)$ are possible  \cite{Armesto:2004ud,Freund:2002ux}, 
the two  given above indicate the higher and lower value of $Q_s$ at
the RHIC energy \cite{Lappi:2007ku} therefore we take these two to set the upper and lower estimate of $Q_s$.
Again using $Q_s/g^2\mu=0.57$ we obtain
$g^2\mu_\mathrm{Pb} = 2$ GeV and $g^2\mu_\mathrm{Pb} = 3$ GeV at the RHIC energy taking respectively the IP-Sat and
naive forms; at the LHC energy $\sqrt{s}=5.02$ TeV the modified GBW fit then leads 
to  $g^2\mu_\mathrm{Pb} = 3.4$ GeV and $g^2\mu_\mathrm{Pb} = 5$ GeV for the two cases.

Before going ahead, we remark that the parameters quoted above (that are widespread in the literature), 
correspond to the color-$SU(3)$ case; on the other hand, in this work we analyze for simplicity the evolution of Glasma
for the color-$SU(2)$ case. We have checked that if we rescale the $g^2\mu$ from $SU(3)$ to $SU(2)$,
we get an effect on the nuclear modification factors of at most the $10\%$:  given this tiny effect of the rescaling,
and in order to avoid confusion with the existing literature, we use the $SU(3)$ parameters here; 
in the Appendix we show one result for the $SU(2)$ parameter set.

The static color sources $\{\rho\}$ generate pure gauge fields which in the forward light cone combine
and give the initial boost-invariant Glasma fields. In order to compute these fields
we  first solve the Poisson equations for the gauge potentials
generated by the color charge distribution of the nuclei $A$ and $B$, namely
\begin{equation}
-\partial_\perp^2 \Lambda^{(A)}(\bm x_T) = \rho^{(A)}(\bm x_T)
\end{equation}
(a similar equation holds for the color charge distribution belonging to $B$). Wilson lines are computed as
$
V^\dagger(\bm x_T) = e^{i \Lambda^{(A)}(\bm x_T)}$,
$W^\dagger(\bm x_T) = e^{i \Lambda^{(B)}(\bm x_T)}$
and the pure gauge fields of the two colliding objects are given by
$
\alpha_i^{(A)} = i V \partial_i V^\dagger$,
$\alpha_i^{(B)} = i W \partial_i W^\dagger$.
In terms of these given fields the solution of the CYM in the forward light cone
at initial time, namely the Glasma gauge potential, are given by
$A_i = \alpha_i^{(A)} + \alpha_i^{(B)}$~ for $i=x,y$ and $A_{\eta} = 0$,
and the initial longitudinal Glasma fields are \cite{Kovner:1995ja,Kovner:1995ts}
\begin{eqnarray}
&& E^{\eta} = i\sum_{i=x,y}\left[\alpha_i^{(B)},\alpha_i^{(A)}\right], \label{eq:f1}\\
&& B^{\eta} = i\left(
\left[\alpha_x^{(B)},\alpha_y^{(A)}\right]  + \left[\alpha_x^{(A)},\alpha_y^{(B)}\right]
\right),\label{eq:f2}
\end{eqnarray}
while the transverse fields are vanishing at initial time.

The dynamical evolution of the fields that we study here is given by the
Classical Yang-Mills (CYM) equations in the case of a box with longitudinal expansion, which are very well known,
see for example \cite{Romatschke:2005pm,Fujii:2008km}. 
The hamiltonian density is given by
\begin{equation}
H = {\rm Tr}\left[\frac{1}{\tau^{2}}E^{i}E^{i}+E^{\eta}E^{\eta}+\frac{1}{\tau^{2}}B^{i}B^{i}+B^{\eta}B^{\eta}\right],
\label{eq:H}
\end{equation}
where the trace is over the adjoint color indices and $\tau$ denotes the proper time.
The equations of motion for the fields and conjugate momenta are
\begin{eqnarray}
  E^{i} &=&  \tau \partial_{\tau} A_{i} ,\\
  E^{\eta} &=&  \frac{1}{\tau} \partial_{\tau} A_{\eta} ,\\
  \partial_{\tau}E^i &=& \frac{1}{\tau}D_\eta F_{\eta i}+\tau D_j F_{ji}  ,\\
\partial_{\tau}E^\eta &=& \frac{1}{\tau}D_j F_{j\eta}.
\end{eqnarray}
These equations are solved in a $4~\mathrm{fm}\times 4~\mathrm{fm}$ box with periodic boundary
conditions in the transverse plane, while rapidity independence is assumed;
the lattice spacing is $a=0.04$ fm .

\section{Charm and beauty in the evolving Glasma}
 
We initialize the momentum distribution of charm and beauty with the prompt distribution obtained
within
Fixed Order + Next-to-Leading Log (FONLL) QCD which
describes the D-mesons and B-meson spectra in $pp$ collisions after fragmentation~\cite{FONLL, Cacciari:2012ny,Cacciari:2015fta}
\begin{equation}
\left.\frac{dN}{d^2 p_T}\right|_\mathrm{prompt} = \frac{x_0}{(1 + x_{3}{p_{T}}^{x_1})^{x_2}};\label{eq:HQ_1}
\end{equation}
the parameters that we use in the calculations are $x_0=20.2837$, $x_1=1.95061$, $x_2=3.13695$ and $x_3=0.0751663$ for 
c quark; $x_0=0.467997$, $x_1=1.83805$, $x_2=3.07569$ and $x_3=0.0301554$ for b quark;
the slope of the distribution has been calibrated to a collision at 5.02 TeV.
Moreover, we assume that the initial longitudinal rapidity share the same value of space-time rapidity $y=\eta$.
Initialization in coordinate space is done as follows:
for the pA collisions we use the gaussian weight in Eq.~\eqref{eq:xi_distri} to distribute heavy quarks in the transverse plane,
while these are distributed with uniform probability in the rapidity direction;
for AA collisions we also distribute heavy quarks with uniform probability in the transverse plane.
Finally,  we assume that the formation proper time
of heavy quark  is $\tau_\mathrm{formation}=1/(2m_{HQ})\approx 0.06$ fm/c for charm quarks having mass $m_c=1.5$ GeV;
for beauty we use $m_b=4.5$ GeV which gives $\tau_\mathrm{formation}\approx 0.02$ fm/c.

The momentum evolution of heavy quarks  in the evolving Glasma is studied by the Wong 
equations  \cite{Wong:1970fu,Boozer}, that for a single heavy quark can be written as
\begin{eqnarray}
&&\frac{d x_i}{dt} = \frac{p_i}{E},\\
&&E\frac{d p_i}{dt} = Q_a F^a_{i\nu}p^\nu,\label{eq:diff1}\\
&&E\frac{d Q_a}{dt} =  Q_c\varepsilon^{cba}  A_{b\mu} p^{\mu},
\end{eqnarray}
where $i=x,y,z$ and $E = \sqrt{\bm p^2 + m^2}$ with $m$ is the mass of the heavy quark.  
In the above equations, the first two equations are the familiar Hamilton equations of motion for the 
coordinate and its conjugate
momentum, while  the third equation corresponds to the gauge invariant color current conservation.
In the third equation $Q_a$ corresponds to the
classical heavy quarks color charge: we initialize this by a uniform distribution with support in the range $(-1,+1)$.
For each $c,b$ quark we produce a $\bar c,\bar b$ quark as well: for this we assume the same
initial position of the companion quark, opposite momentum and a new random color charge
(color singlet condition of the produced quark-antiquark pair is not a requirement because most heavy quarks are produced via gluon fusion).
Solving the Wong equations is equivalent to solve the Boltzmann-Vlasov equations for the heavy quarks 
in a collisionless plasma, which propagate in the evolving Glasma.
In fact, the Boltzmann-Vlasov equation can be solved by means of the test particle method which amounts
to solve the classical equations of motion of the particles in the background of the evolving gluon field.

We neglect the effect of the color current carried by the heavy quarks on the gluon field:
this approximation is usually used to study the propagation of heavy quarks in a thermal QGP bath
and sounds quite reasonable due to
the small number of heavy quarks produced by the collision, as well as due to their large mass,
both of these factors eventually leading to a negligible color current density.
The term on the right hand side of Eq.~\eqref{eq:diff1} is responsible of diffusion in momentum 
space  \cite{Ruggieri:2018rzi,Sun:2019fud,Ruggieri:2019zos};
besides,  we should add a dissipative term,
$-E\Gamma p_i$,
that takes into account the energy loss due to gluon radiation.
We will estimate roughly the effect of gluon radiation in Section V in the case of charm quark only,
by firstly computing the diffusion coefficient
then using the Fluctuation-Dissipation Theorem to compute the drag which describes the energy loss.
We anticipate here that, within our approximations, the effect of the drag is not substantial for the
range of $p_T$ that we consider in our study unless the average effective temperature of the gluon bath
is very low.

Whether a quark-gluon plasma is formed or not in pA collisions is still an open question.
For the pA collisions we assume that all the dynamics can be described
by the interaction with the evolving Glasma fields, without the formation of a quark-gluon plasma;
eventually heavy quarks hadronize by fragmentation.
At the end of this evolution we adopt a standard fragmentation for the heavy
quark to D/B-meson~\cite{Pet}, with
\be
f(z) \propto
\frac{1}{z \left( 1- \frac{1}{z}- \frac{\epsilon_c/b}{1-z} \right)^2}
\label{fg}
\ee
where $z=p_D/p_c (p_B/p_b)$ is the momentum fraction of the D(B)-meson fragmented from the heavy quark and
$\epsilon_c$($\epsilon_b$) is a free parameter to fix the shape of the fragmentation function in order to
reproduce the D/B-meson production in $pp$ collisions~\cite{Scardina:2017ipo}.
On the other hand, for AA collisions we do not perform any fragmentation since we can describe only
the evolution in the early stage of the collision: our result for the spectrum should eventually be fed to 
a relativistic transport code as an initialization for heavy quarks in the quark-gluon plasma phase
as done in \cite{Sun:2019fud}.

\section{Results}

In this section we present our results. We firstly show the color fields, the energy density
and define an effective average temperature of the evolving gluon fields. Then we show 
the $R_\mathrm{pPb}$ for $D$ and $B$ mesons, and the $R_\mathrm{PbPb}$ for charm and beauty quarks.

\subsection{Evolving color fields}
\begin{figure}[t!]
\begin{center}
\includegraphics[width=0.45\textwidth]{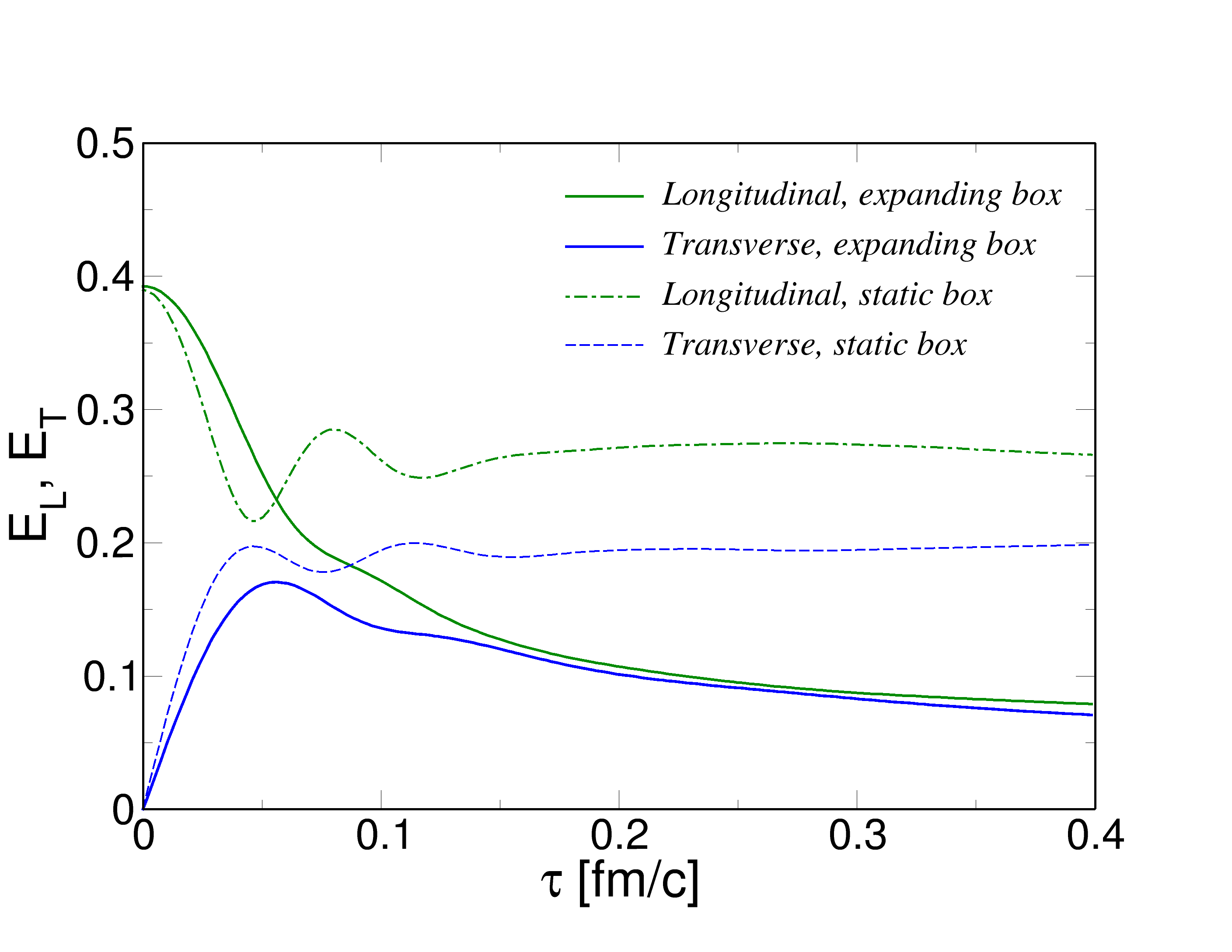}
\end{center}
\caption{\label{Fig1a}
Color electric fields, measured in lattice units, as a function of proper time for the expanding box (solid lines) and the static box (dashed and dot-dashed lines).}
\end{figure}

In Fig.~\ref{Fig1a} we plot the color electric field, measured in lattice units, as a function of proper time;
the behavior of the color magnetic field is similar so we do not show it.
In the figure, 
solid green line corresponds to the longitudinal field while the solid green thick line denotes the averaged transverse field, namely $E_T = \sqrt{E_x^2 + E_y^2}/2$.
In all cases, the fields have been averaged over the box.
For comparison, we also show the fields in the case of the static box.
The calculations correspond to the set-up for a Pb-Pb collision and $g^2\mu=3.4$ GeV.
We notice that regardless of the longitudinal expansion, $E_T$ forms within $0.1$ fm/c in proper time, and initially its magnitude
is comparable with that of the static box. At larger proper times, the magnitude of both longitudinal and transverse fields decreases due to the
expansion that dilutes the energy density; nevertheless, up to $\tau=0.4$ fm/c the size of $E_T$ is still substantial, being $\approx 40\%$ of the analogous field 
obtained for the static box. This $E_T$ causes the diffusion of the transverse momentum of heavy quarks.

\begin{figure}[t!]
\begin{center}
\includegraphics[width=0.45\textwidth]{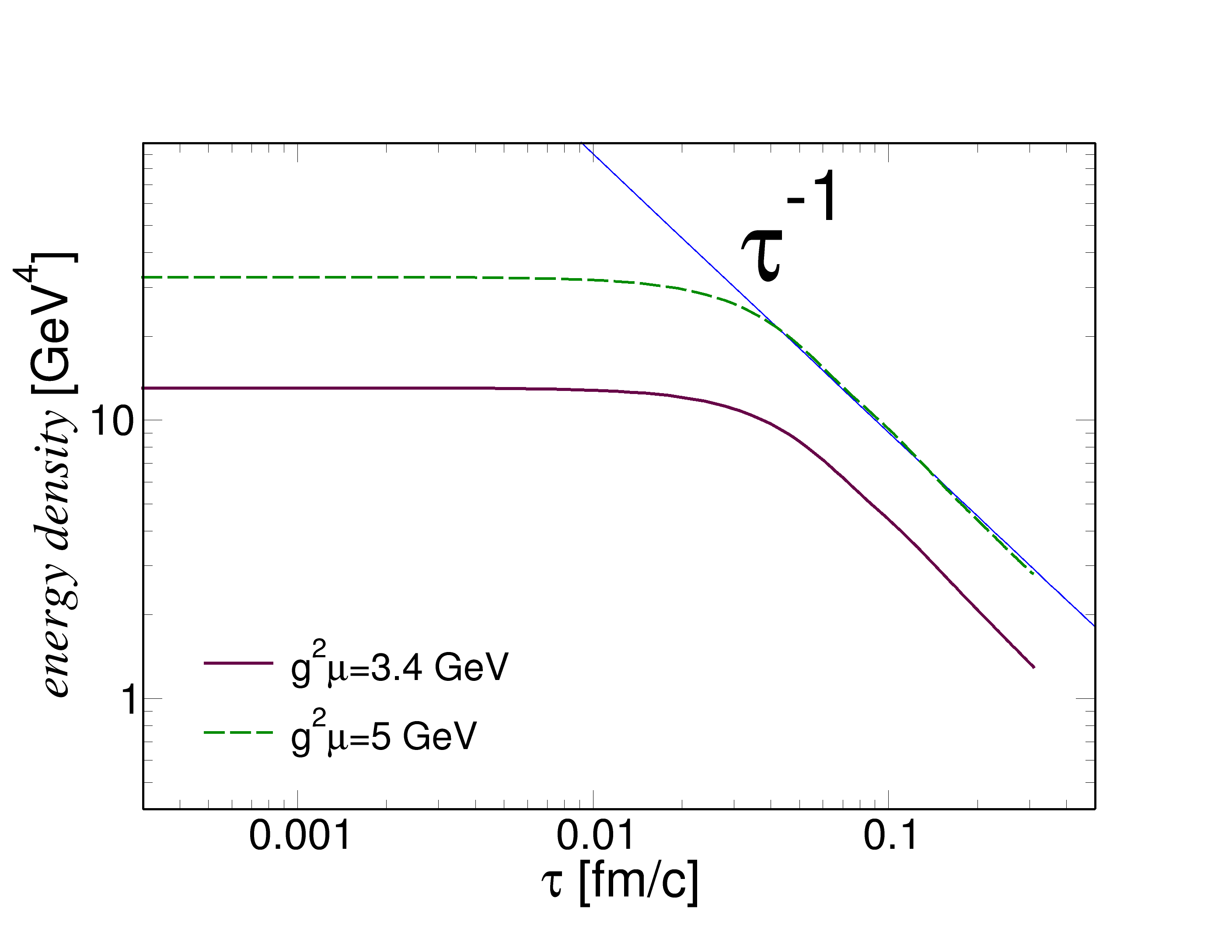}
\end{center}
\caption{\label{Fig:energy_density}
Average energy density of the gluon field versus proper time,
for a Pb-Pb collision with $g^2\mu=3.4$ GeV (solid maroon line) and $g^2\mu=5$ GeV
(dashed green line). 
Blue thin line corresponds to $\tau^{-1}$ which we draw to emphasize the free streaming
that develops for $\tau\gtrsim 0.05$ fm/c.}
\end{figure}

In Fig.~\ref{Fig:energy_density} we plot the average energy density 
of the evolving gluon field, $\rho$, as a function of the proper time,
computed for a Pb-Pb collision with $g^2\mu_\mathrm{Pb}=3.4$ GeV
(solid maroon line) and $g^2\mu=5$ GeV
(dashed green line). 
In the figure we have also drawn the blue thin line that corresponds to $\tau^{-1}$:
it emphasizes the free streaming
that develops for $\tau\gtrsim 0.05$ fm/c, corresponding to a vanishing longitudinal pressure \cite{Fukushima:2011nq,Ruggieri:2017ioa}.

\begin{figure}[t!]
\begin{center}
\includegraphics[width=0.45\textwidth]{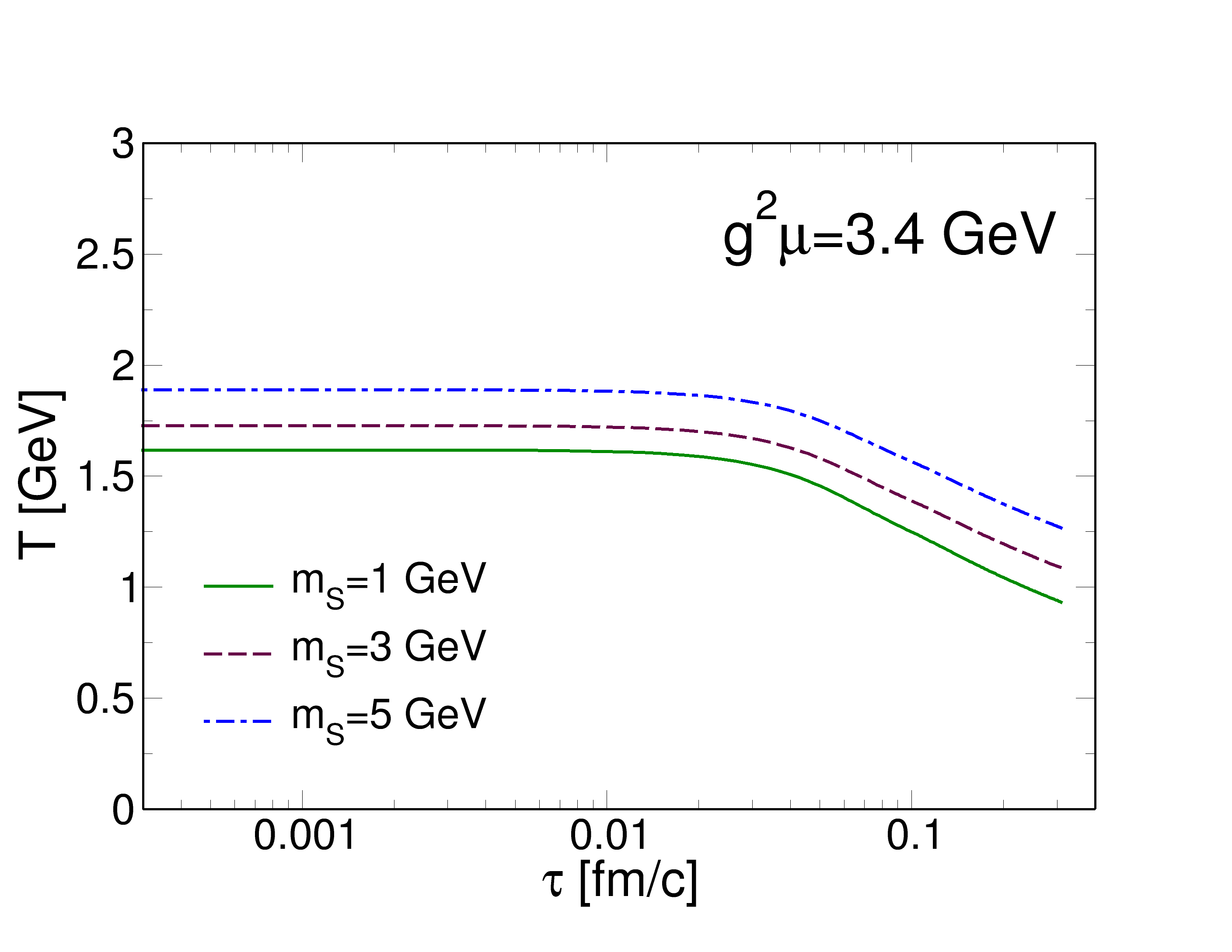}\\
\includegraphics[width=0.45\textwidth]{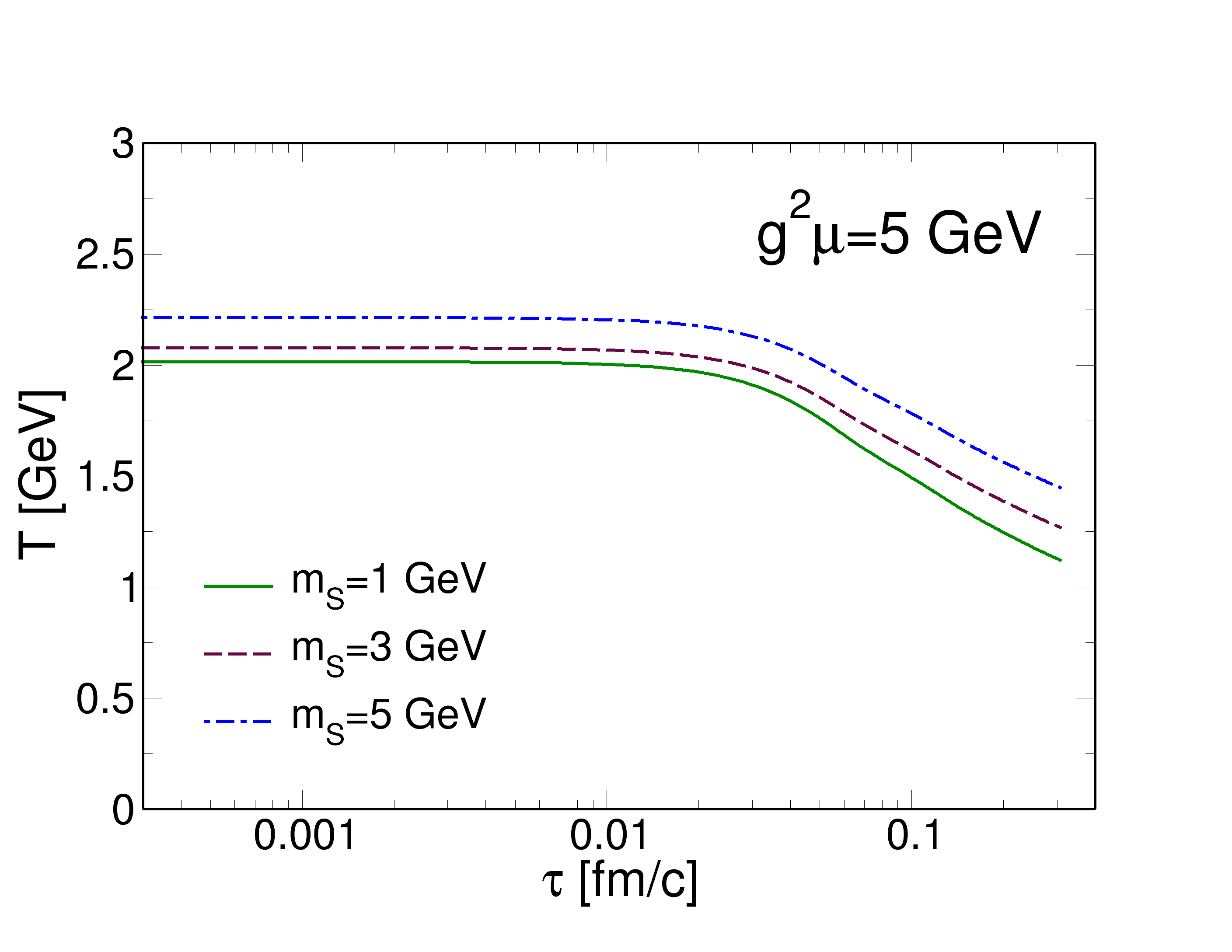}
\end{center}
\caption{\label{Fig:energy_density_temp}
{\em Upper panel.} Effective temperature versus proper time, computed
for a Pb-Pb collision with $g^2\mu=3.4$ GeV and for 
$m_S=1$ GeV (green solid line), $m_S=3$ GeV (maroon dashed line),  
$m_S=5$ GeV (blue dot-dashed line).
{\em Lower panel.} Same as upper panel, with $g^2\mu=5$ GeV.}
\end{figure}

For what we will discuss in the next sections, it is useful to define an effective temperature from the energy density,
assuming the evolving gluon system can be represented as an ensemble of thermalized
gluons with average temperature $T$ and screening mass $m_S$, namely 
\begin{equation}
\rho = 2(N_c^2-1)\int\frac{d^3p}{(2\pi)^3}\frac{\sqrt{p^2+m_S^2}}{e^{\sqrt{p^2+m_S^2}/T}-1},
\label{eq:rho_gluons_T}
\end{equation}
with $N_c$ the number of colors. From the results shown in Fig.~\ref{Fig:energy_density} we extract
the effective temperature for three values of $m_S$, and we show the results in 
Fig.~\ref{Fig:energy_density_temp} for two values of $g^2\mu$:
the green solid lines correspond to $m_S=1$ GeV, the maroon dashed lines to $m_S=3$ GeV
and the blue dot-dashed lines to $m_S=5$ GeV.
The screening mass can be read from the gauge invariant correlators \cite{Ruggieri:2017ioa}
and it has been found to be of the order of $g^2\mu$: we consider here three values of $m_S$ to check
how the predictions depend on $m_S$.  
The average temperature is pretty high at initial time, but it reaches $T\approx 1$ GeV
within $\tau\approx 0.4$ fm/c for the lowest value of $m_S$,
being a bit higher for larger values of $m_S$. Clearly chosing a smaller value of $m_S$ would give an effective
temperature which matches the initialization temperature of hydro and relativistic transport calculations,
$T_\mathrm{in}\approx 0.6$ GeV.

\subsection{Nuclear modification factor for p-Pb collisions}

\begin{figure}[t!]
\begin{center}
\includegraphics[width=0.45\textwidth]{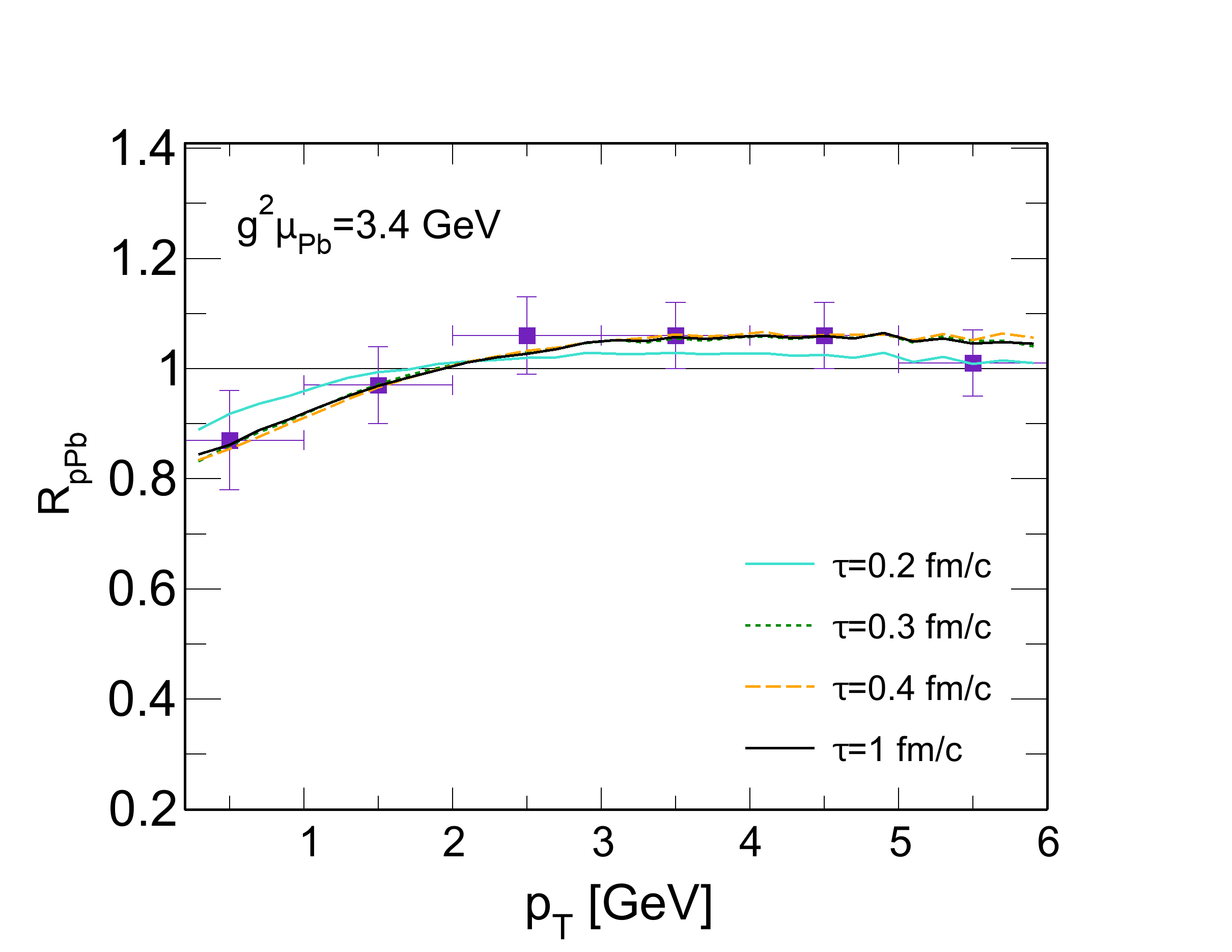}\\
\includegraphics[width=0.45\textwidth]{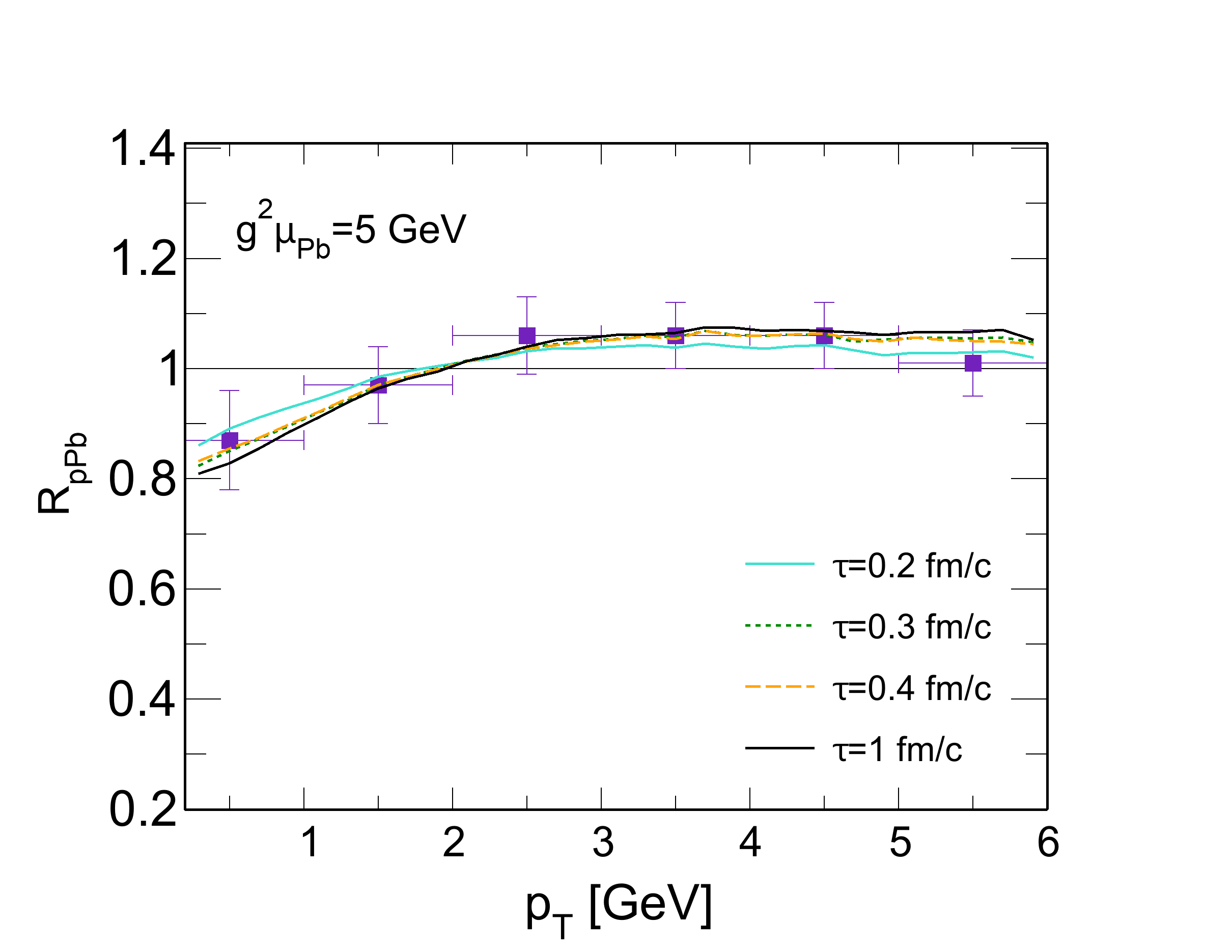}
\end{center}
\caption{\label{Fig1}
Nuclear suppression factor($\left | y \right |<0.2$) for the $D-$ meson versus $p_T$ at different times.
Upper panel corresponds to $g^2\mu_\mathrm{Pb} = 3.4$ GeV while in the lower panel $g^2\mu_\mathrm{Pb}=5$ GeV. 
Experimental data correspond to the backward rapidity side 
of the LHCb collaboration~\cite{Aaij:2017gcy}.}
\end{figure}

In Fig.~\ref{Fig1}, we plot the D-meson nuclear suppression factor, $R_\mathrm{pPb}$,  at different time for $D-$meson
and for two different values of $g^2\mu$ in the Pb nucleus, namely 
$g^2\mu = 3.4$ GeV (upper panel) and $g^2\mu = 5$ GeV (lower panel).
The last is a large value of $g^2 \mu$, but still within standard estimates. 
We  notice that anyway the effect on $R_\mathrm{pPb}$ only mildly depends on the $g^2\mu$.
The evolution of charm has been studied via the Wong equations, then at the time shown in the figure we 
have adopted the standard fragmentation procedure to get the $D-$meson spectra.
We observe that the $R_\mathrm{pPb}$ substantial  deviation  from one because of the interaction of the
charm with the gluon fields. 
We notice that due to the heavy quark  interaction with the gluon fields,  low momentum charm quarks are shifted
to high momentum. This effect was first observed in \cite{Ruggieri:2018rzi} and
named as the Cathode Tube Effect. Due to Cathode tube effect we observe a enhancement of the
$R_\mathrm{pPb}$ at intermediate $p_T$ associated to a depletion, a consequence of heavy quark
number conservation, at low momenta. 
In comparison with \cite{Ruggieri:2018rzi},
in this present work we have studied the effect in a more
realistic scenario including the longitudinal expansion. 
In Fig.~\ref{Fig1}, we consider
$g^2\mu = 3.4$ GeV for the Pb nucleus. Although there is uncertainty on the life time of Glasma
produced in the p-Pb collisions, we observe that the nuclear suppression factor stops evolving
at $\tau\approx 0.3$ fm/c: this is mainly due to the
expansion which was  missing in the earlier calculations~\cite{Ruggieri:2018rzi}.
Such a property of self-quenching in a short time scale is likely to favor a 
smooth transition into the subsequent evolution of the system in the QGP.
In the lower panel of Fig.~\ref{Fig1}, we plot the D-meson nuclear suppression factor in Glasma corresponding to
$g^2\mu_\mathrm{Pb}=5$ GeV. Larger value of $g^2\mu_\mathrm{Pb}$ corresponds to more color charges,
hence, large energy density. This leads to the larger enhancement at intermediate momentum.
The fact that the  shape of our  $R_\mathrm{pPb}$
can reproduce that measured in experiments,  might suggest that at least part of the measured $R_\mathrm{pPb}$
comes from the propagation of the charm quarks in the Glasma, and might be considered as the signature of the
Glasma itself. 

The net effect that we find on charm spectra seems different from what is usually observed 
in a quark-gluon plasma medium, in which the nuclear modification factor is larger than one at low $p_T$
then it decreases monotonically smaller at larger $p_T$. 
However, it should be considered that in the early stage the energy density of the gluon bath is very large,
therefore it is natural that low-$p_T$ heavy quarks diffuse to higher $p_T$
and energy loss plays a minor role here.
We will show in the next section that including a term that describes energy loss by radiation,
the main effect will be still the diffusion from low to high $p_T$ unless we keep the average energy density
of the gluon bath very low in the full evolution, comparable with that of the quark-gluon plasma
at the initialization time.
This is in agreement with~\cite{Ruggieri:2019zos} where it has been shown 
that assuming a Fluctuation Dissipation Theorem,
the diffusion of the heavy particle dominates over the energy loss if the initial momentum
of the heavy particle is smaller than the average energy of the bulk.

\subsection{Nuclear modification factor for Pb-Pb collisions}
\begin{figure}[t!]
\begin{center}
\includegraphics[width=0.45\textwidth]{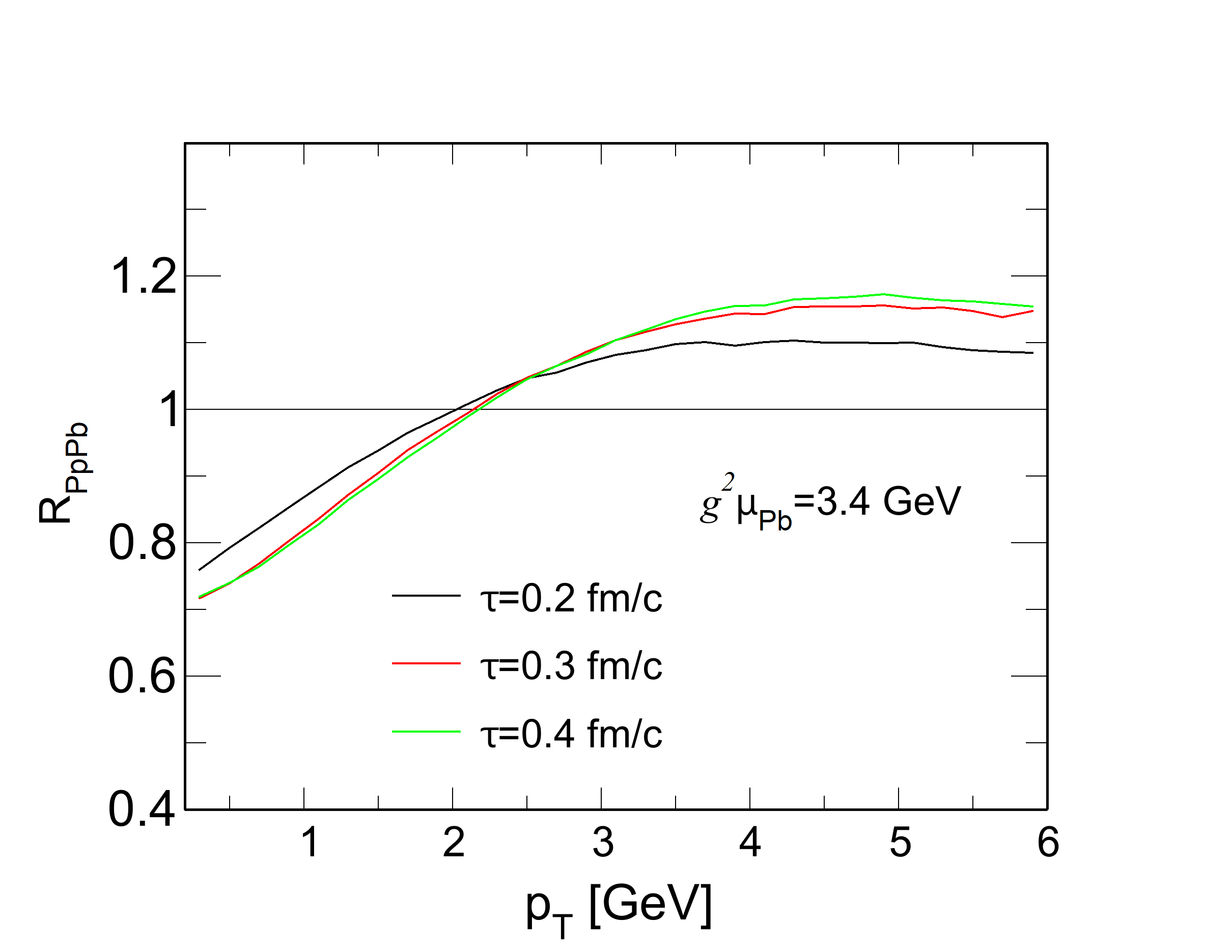}\\
\includegraphics[width=0.45\textwidth]{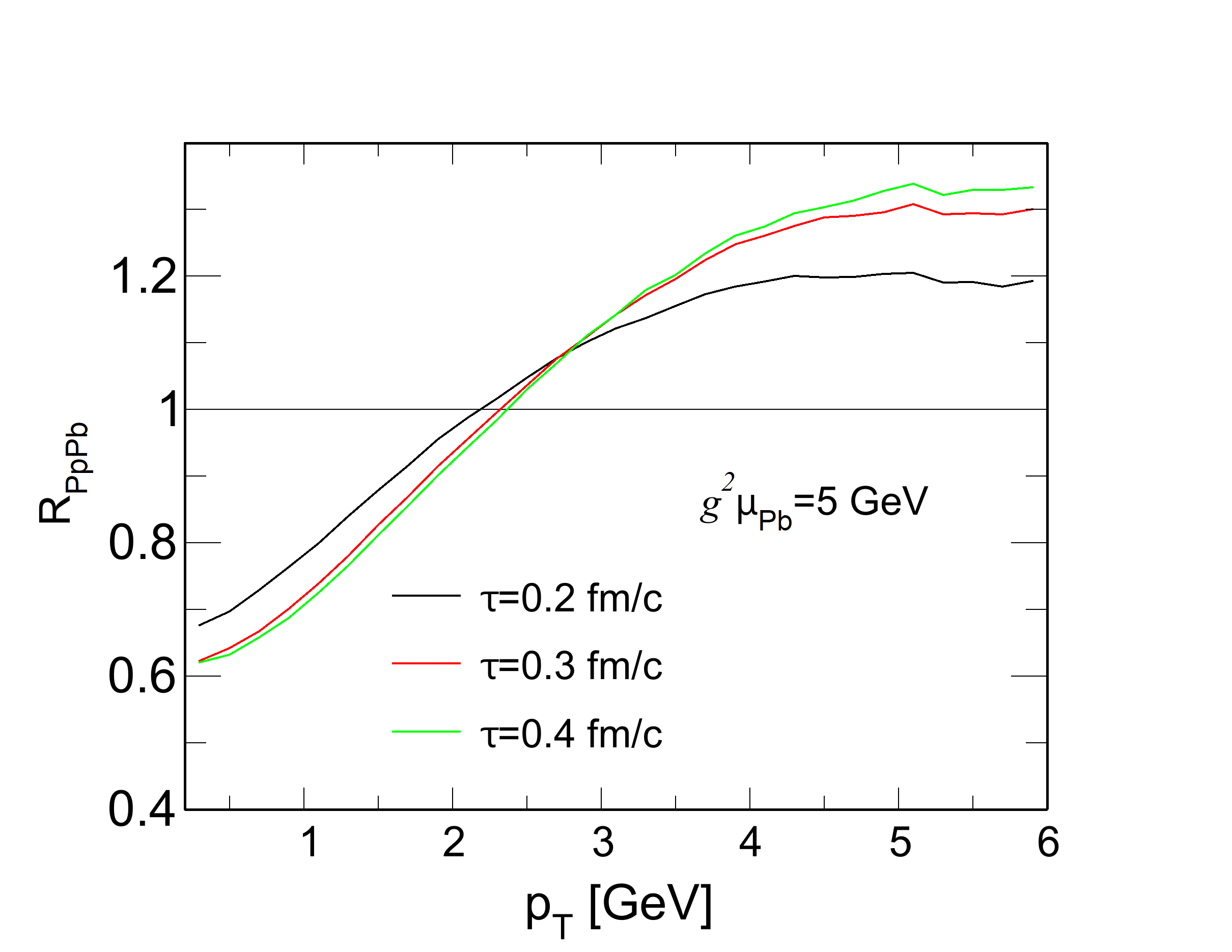}
\end{center}
\caption{\label{Fig2}
Nuclear suppression factor of charm versus $p_T$ for a Pb-Pb collision, at different times.
Upper panel corresponds to $g^2\mu_\mathrm{Pb} = 3.4$ GeV while in the lower panel $g^2\mu_\mathrm{Pb}=5$ GeV.}
\end{figure}

In nucleus-nucleus collisions the Glasma phase can act as the pre-equilibrium phase before the QGP phase.
Recently the impact of initial static Glasma phase on heavy quark dynamics has been reported in~\cite{Sun:2019fud}.
It is observed
that the dynamics in evolving Glasma stage  leads to a larger final elliptic flow ($v_2$) inducing a relation between
$R_{AA}$ and $v_2$ that is quite close to the experimental measurements. Keeping its significance in mind, we
extended our calculation to nucleus-nucleus collisions.  
In Fig.~\ref{Fig2}, we plot  $R_\mathrm{PbPb}$ of charm in evolving Glasma for
$g^2\mu_\mathrm{Pb}=3.4$ GeV (upper panel) and $g^2\mu_\mathrm{Pb}=5$ GeV (lower panel)
at different proper times.
The effect on the charm spectrum is much stronger in nucleus-nucleus
collisions in comparison with p-nucleus collisions: this is due to the larger energy density produced in
nucleus-nucleus collisions than p-nucleus collisions. 
After the evolution in the initial gluon fields,
the quark-gluon plasma phase should be considered 
and our result should serve as the initial condition of the charm spectrum for the QGP
phase evolution. 
Once again we observe a saturation of the $R_\mathrm{PbPb}$ at $\tau\approx 0.3$ fm/c. 
It is remarkable that although expansion is present, the impact of initial evolution in the gluon field
is still quite substantial in comparison
with the earlier results in which expansion was not considered~\cite{Sun:2019fud}.

\begin{figure}[t!]
\begin{center}
\includegraphics[width=0.45\textwidth]{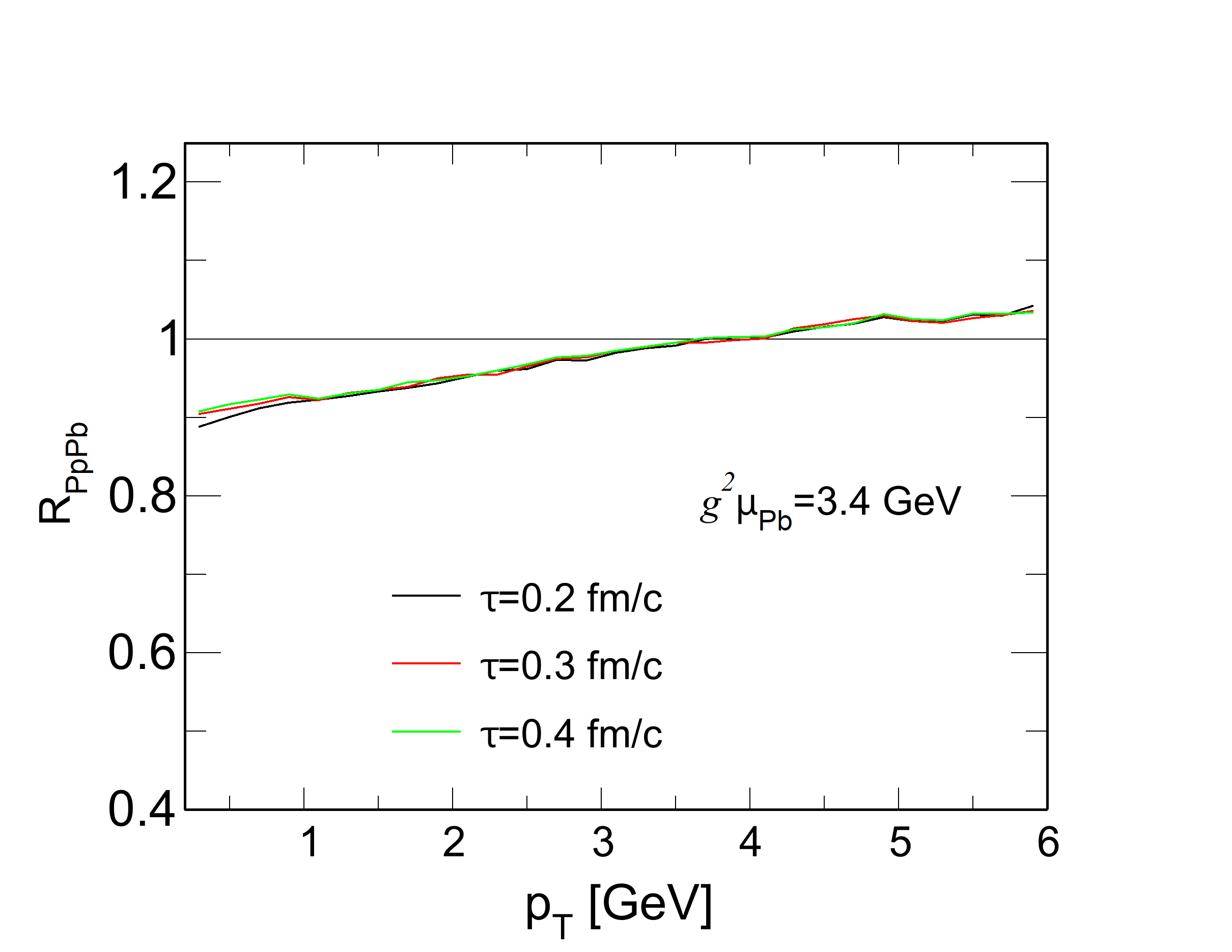}\\
\includegraphics[width=0.45\textwidth]{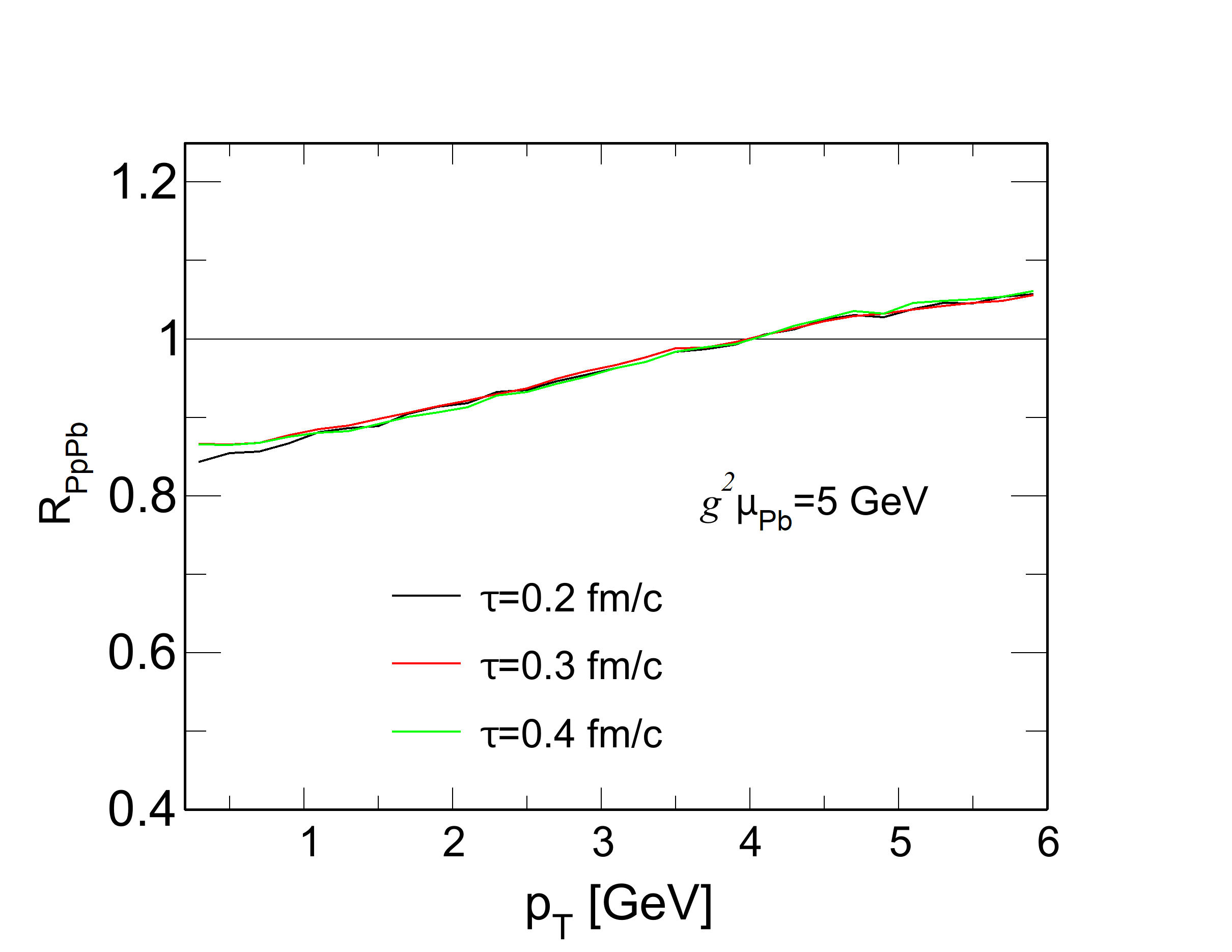}
\end{center}
\caption{\label{Fig5}
Nuclear suppression factor of beauty versus $p_T$ for a Pb-Pb collision, at different times.
Upper panel corresponds to $g^2\mu_\mathrm{Pb} = 3.4$ GeV while in the lower panel $g^2\mu_\mathrm{Pb}=5$ GeV.}
\end{figure}

We further extend our model to study the evolution of beauty. 
In Fig.~\ref{Fig5}, we present the $R_\mathrm{PbPb}$ of beauty
for the aforementioned two values of $g^2\mu_\mathrm{Pb}$.
The trend we observe for beauty is the same we have found for charm:
quantitatively the effect on the spectrum of beauty is much milder,
of course because of the larger mass of this quark compared to that of charm.
Overall, 
the effect on beauty is still significant 
and might impact the later
dynamics in the QGP phase \cite{Sun:2019fud}.  

\section{Estimate of the radiation reaction}
In the previous Section we have shown the results obtained neglecting the energy loss of the heavy quarks:
this can be modeled by modifying Eq.~\eqref{eq:diff1} as
\begin{equation}
E\frac{d p_i}{dt} = Q_a F^a_{i\nu}p^\nu - E\Gamma p_i,\label{eq:diff2}
\end{equation}
where $\Gamma$ is the drag coefficient.
 In this Section we give a rough estimate of the radiation reaction, then evaluate its effect
on the nuclear modification factors. We limit ourselves to show the results obtained for the charm quark since those for the beauty quark are similar.
Our strategy is to firstly evaluate an average diffusion coefficient for the transverse momentum, $D$, of charm, defined via the equation
\begin{equation}
 \sigma^2\equiv \left \langle(p_{T}-\langle p_{T}\rangle)^2\right \rangle =2 D \tau~+~\mathrm{constant}, \label{eq:FD_222}
\end{equation}
where the average is taken with the charm spectrum,
then assume that the Fluctuation-Dissipation theorem (FDT),
\begin{equation}
  D= \Gamma  E T, \label{eq:FD_333}
\end{equation}
relates $D$ to $\Gamma$ with $E$ corresponding to the kinetic energy of the heavy quark. 
A similar assumption between drag and diffusion coefficients, although via a more rigorous implementation,
has been already adopted in \cite{Mrowczynski:2017kso}.
One major drawback of this assumption is that
the FDT is rigorously valid only for a bath in thermal equilibrium, which is a condition that is not satisfied
by the evolving gluon field considered here.
However, the calculation of $\Gamma$ without a FD theorem would require the calculation of the
gluon radiation from quarks propagating in a random gluon field, which is beyond the
scope of this study. Therefore, we follow \cite{Mrowczynski:2017kso} and we chose to
estimate the drag coefficient by Eq.~\eqref{eq:FD_333} and replacing $T$
with the average temperature extracted from the energy density, see 
Fig.~\ref{Fig:energy_density} and Eq.~\eqref{eq:rho_gluons_T}.

\begin{figure}[t!]
\begin{center}
\includegraphics[width=0.45\textwidth]{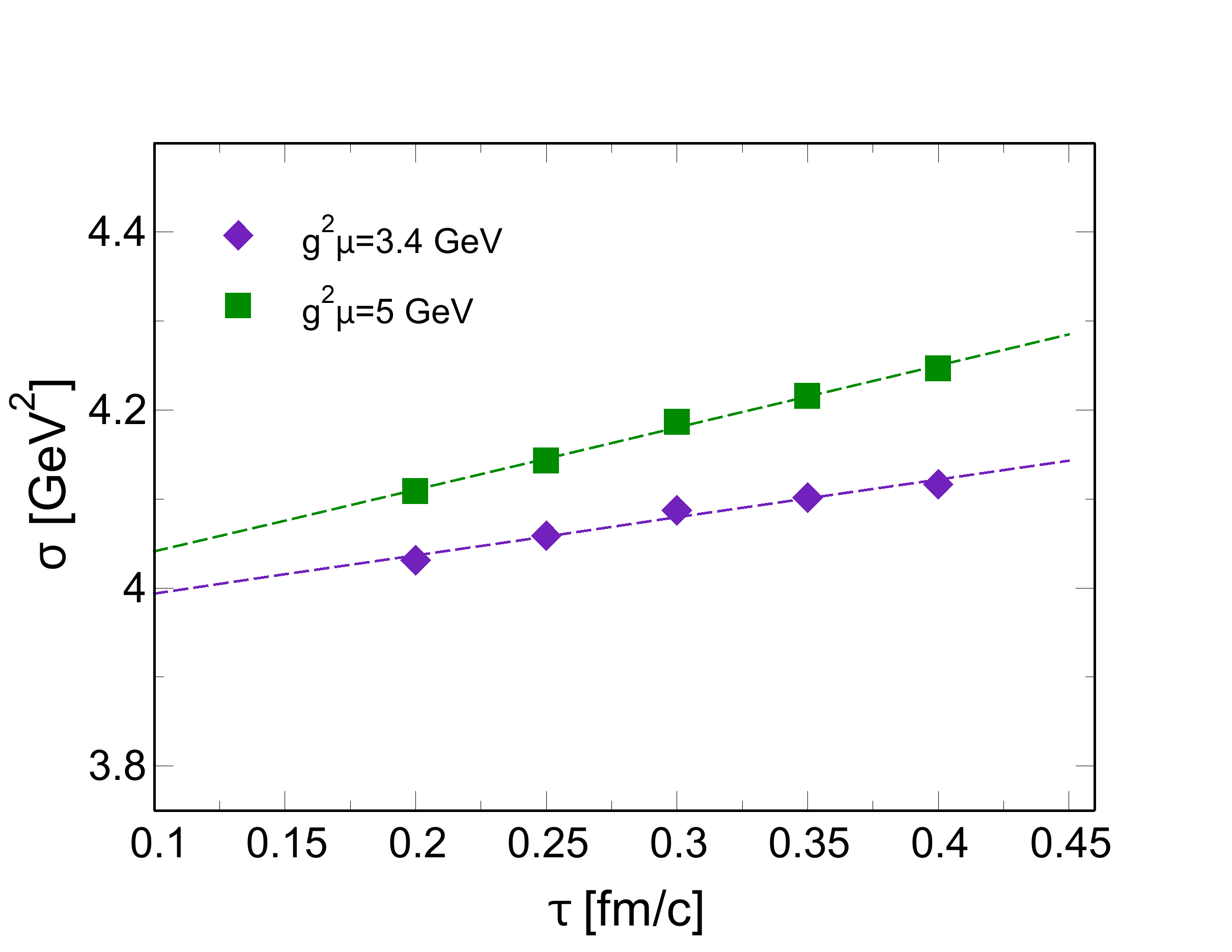}
\end{center}
\caption{\label{FigGGG}
Transverse momentum variance as a function of proper time, for a Pb-Pb collision. 
Indigo diamonds correspond to $g^2\mu = 3.4$ GeV while green squares to $g^2\mu=5$ GeV.
Dashed lines denote the linear fit Eq.~\eqref{eq:FD_222}.}
\end{figure}

In Fig.~\ref{FigGGG} we plot the variance of the transverse momentum for the setup of a Pb-Pb collision.
The result has been obtained by solving the equations of motion of charm in the evolving gluonic background, Eq.~\eqref{eq:diff1},
which takes into account the diffusion only. In the figure, the green squares correspond to our calculation
 for $g^2\mu = 5$ GeV while the indigo diamonds denote the results for $g^2\mu=3.4$ GeV; the dashed lines
correspond to the linear fit Eq.~\eqref{eq:FD_222}.
We use Eq.~\eqref{eq:FD_222} to fit the data in Fig.~\ref{FigGGG} and we find
$D = 0.21~\mathrm{GeV}^2/\mathrm{fm}$ for $g^2\mu =3.4$ GeV, while 
$D = 0.35~\mathrm{GeV}^2/\mathrm{fm}$ for $g^2\mu = 5$ GeV.
We notice that these numbers are in the same ballpark of the pQCD estimates for an average $p_T\approx 1.5-2$ GeV
in the temperature range $0.4-1$ GeV \cite{Das:2013kea}. 
Using these results in Eq.~\eqref{eq:FD_333} we estimate $\Gamma$ for each charm and use it in Eq.~\eqref{eq:diff2}.

\begin{figure}[t!]
\begin{center}
\includegraphics[width=0.45\textwidth]{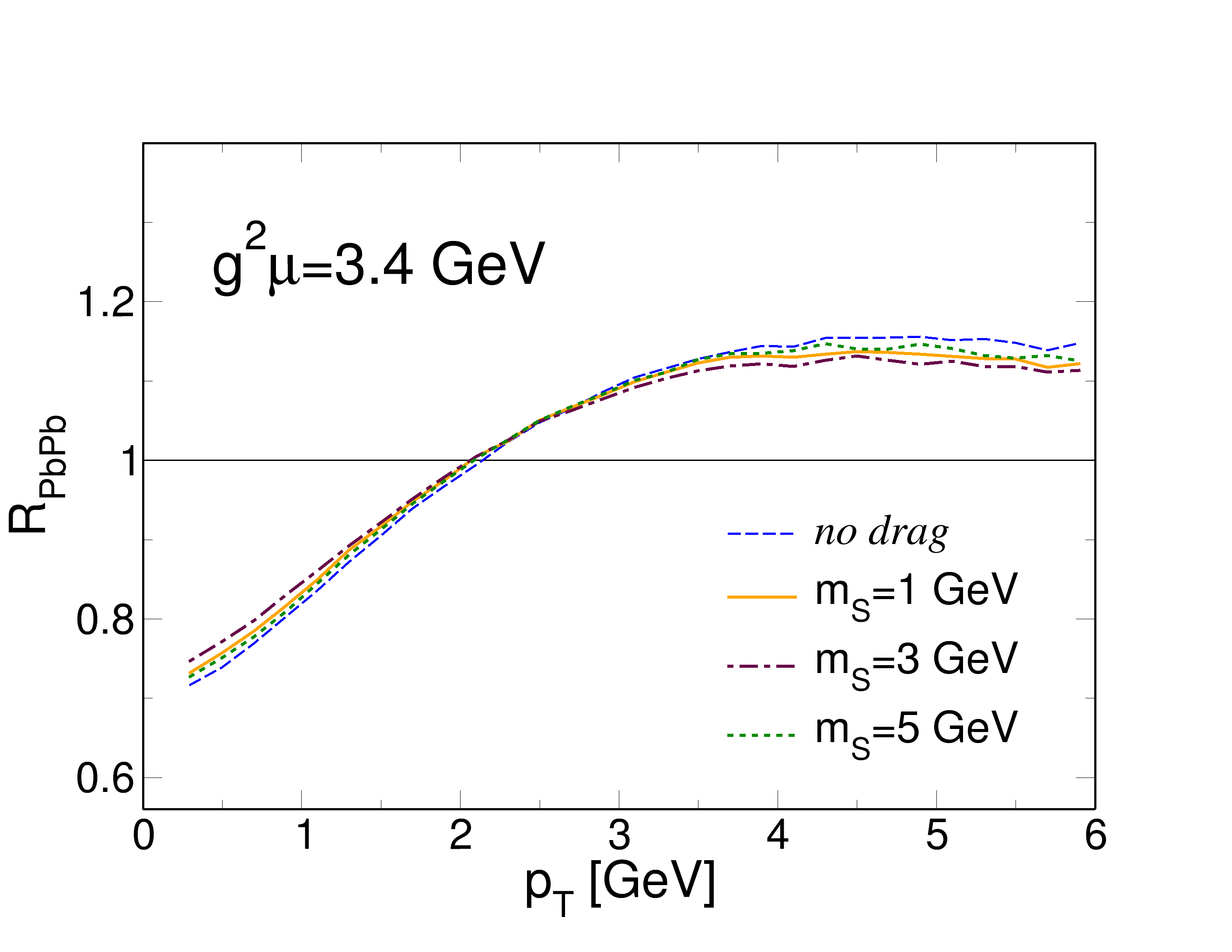}\\
\includegraphics[width=0.45\textwidth]{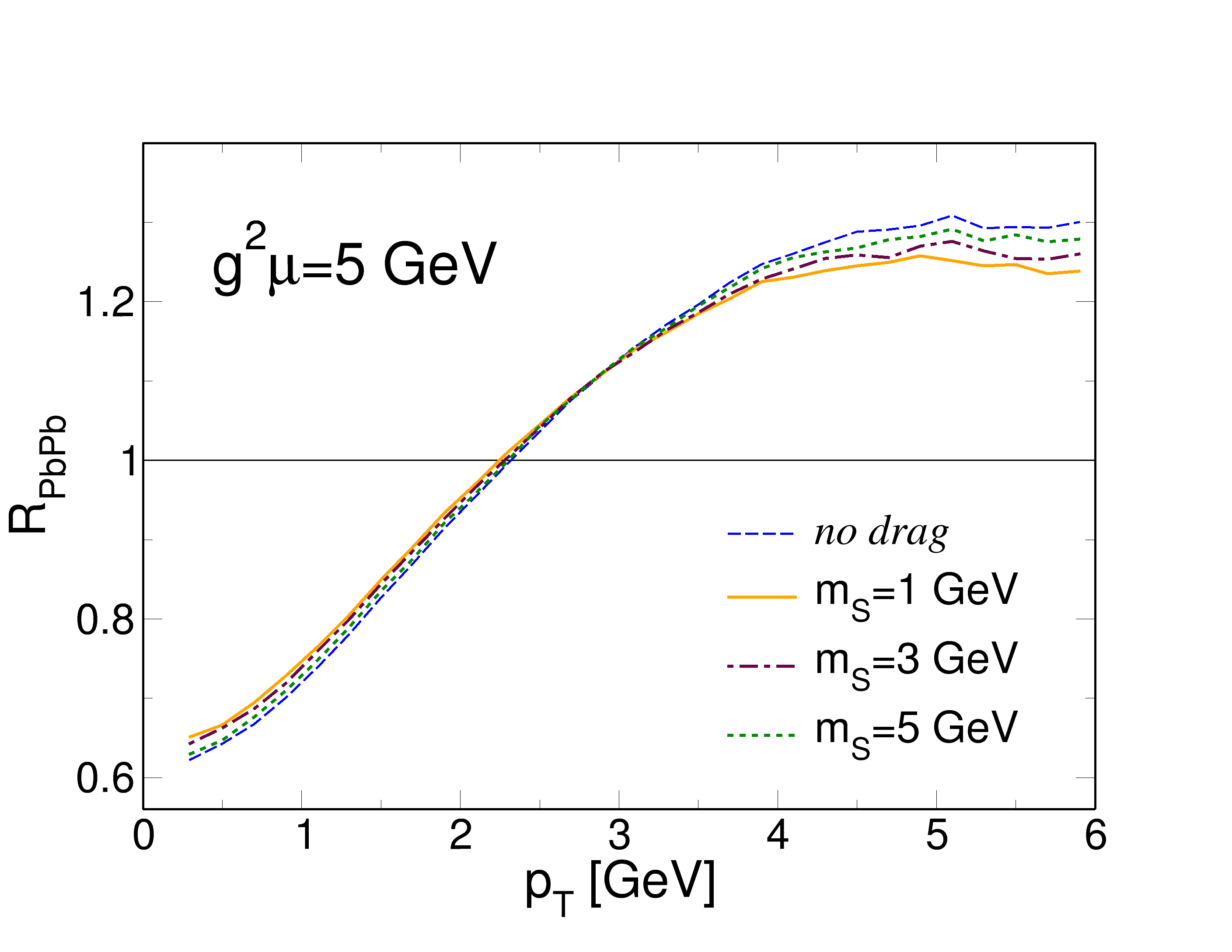}
\end{center}
\caption{\label{FigGGGh}
Comparison of the nuclear modification factor for the set up of Pb-Pb collisions,
obtained neglecting and considering the radiation reaction of the
charm quark. Upper panel corresponds to $g^2\mu=3.4$ GeV and
three screening masses: $m_S=1$ GeV (solid orange line), $m_S=3$ GeV 
(maroon dot-dashed line) and $m_S=5$ GeV (dotted green line). For comparison we also show the result
without drag, see the blue dashed line. Lower panel shows the results
$g^2\mu=5$ GeV and the same values of the screening masses.}
\end{figure}

In Fig.~\ref{FigGGGh} we plot the nuclear modification factor computed
with the drag force estimated by the FDT, using three different screening masses
of the bath. For comparison we also show $R_\mathrm{PbPb}$ computed without taking the drag into account,
see the blue dashed line. Upper panel corresponds to $g^2\mu=3.4$ GeV,
lower panel to $g^2\mu=5$ GeV. In both panels we show the results obtained for
$m_S=1$ GeV (solid orange line), $m_S=3$ GeV 
(maroon dot-dashed line) and $m_S=5$ GeV (dotted green line). 
As expected, the smaller the average $T$ is, the larger the effect of the drag.
On the one hand, these results confirm that as long as the energy density of the gluon medium
remains large in the initial stage, the drag force can be neglected for low-$p_T$ heavy quarks;
on the other hand, when the energy density becomes smaller the drag might be important so energy loss
by gluon radiation might play a role even in the very early stage of the collision. This last circumstance
pushes us to study the problem of gluon radiation in more detail, and we leave a detailed study to a future project.

\section{Conclusions and outlook}
We have studied the diffusion of charm and beauty quarks by means of the Wong
equations in the evolving strong gluon fields produced in the early stage of high energy nuclear collisions.
For the sake of computational simplicity we have addressed the problem in the color-$SU(2)$ case
Initialization of the gluon field has been achieved via the Glasma, carrying longitudinal color-electric
and color-magnetic fields thus neglecting fluctuations. 
For the heavy quark initialization we have considered the standard FONLL
perturbative production tuned in order to reproduce the $D$ and $B$ meson spectrum in proton-proton collisions.
We have set up the saturation scale for both the proton and the Pb nucleus in order to reproduce the expected
one at $\sqrt{s}=5.02$ TeV collisions.
The novelties in comparison with previous works \cite{Ruggieri:2018rzi,Sun:2019fud} are that have added the
longitudinal expansion of the gluon medium (in \cite{Ruggieri:2018rzi} we studied the diffusion in a static box),
and performed a systematic comparative study of pA and AA collisions that is lacking in the literature.
In particular, we have focused on the nuclear modification factors, $R_\mathrm{pA}$ and $R_\mathrm{AA}$,
showing how this is affected by the interaction of the heavy quarks with the gluon medium.
We have also roughly estimated the effect of the radiation reaction on $R_\mathrm{AA}$,
by using a Fluctuation-Dissipation-Theorem-like relation between the drag and the diffusion coefficient, 
leaving a complete study to a future project.
We have found that the spectrum of  charm and beauty is tilted towards higher $p_T$ because of the
interaction with the gluon fields in the early stage: this effect was named the Cathode Tube in \cite{Ruggieri:2018rzi}.

We have firstly computed $R_\mathrm{pPb}$.
The main effect of the interaction of the charm quarks with the gluon field is to shift
the low $p_T$ charm quarks to high $p_T$ states. This ensured a enhancement of the charm quarks yields in
the intermediate $p_T$.
We have found that the  shape of our $R_\mathrm{pPb}$ able to reproduce  that measured by the LHCb collaboration
on the proton side where the effect of shadowing is marginal.  Since in our calculation the shape comes
directly from the propagation of the charm quarks in the evolving Glasma fields, we suggest that at least part of
the measured $R_\mathrm{pPb}$ is the signature of the Glasma formed in high energy collisions.
We have repeated the calculation for the beauty quarks, finding qualitatively similar results to those of 
the charm quarks. We have then studied $R_\mathrm{PbPb}$ for both charm and beauty.
We have found that even when the longitudinal expansion is included,
the impact of the early stage
is substantial both for charm and beauty. 
It is not possible to compare $R_\mathrm{PbPb}$ coming from the early stage with experimental data, since the late evolution 
in the quark-gluon plasma cannot be neglected.
However, even if the evolution in the gluon fields happens at very early times,
this can be related to the observables in Pb-Pb collisions~\cite{Sun:2019fud}. 

In most of the present work we have neglected the effect of gluon radiation on the motion of the heavy quarks.
The inclusion of this is a highly nontrivial step, due to the fact that any estimate of the radiation based on pQCD
should take into account that charm and beauty evolve in a random gluon field
and that gluons are screened with a screening mass $m_D\sim g^2\mu$.
Instead of attacking this problem from the pQCD point of view, we have used a relation
inspired by the Fluctuation-Dissipation Theorem to relate the drag force responsible of the energy loss,
to an averaged diffusion coefficient. The numerical calculation of the latter presents no difficulty within our approach
since it is possible to follow the evolution of $\langle(p_T - \langle p_T\rangle)^2\rangle$ of heavy quarks with time,
thus allowing us to define a diffusion coefficient for the transverse momentum.
We have computed the drag coefficient assuming the evolving gluon bath is described by a gas of massive
gluons thermalized at an average temperature, $T$, which can be estimated within our code from 
the average energy density. 
We have found that the effect of the drag coefficient, that was neglected in previous 
calculations \cite{Ruggieri:2018rzi,Sun:2019fud}, is certainly present but does not seem enough to cancel the Cathode Tube effect
as long as the average temperature remains $T\approx 1$ GeV,
as already anticipated in \cite{Ruggieri:2019zos}.
These results are encouraging since they show that neglecting energy loss, we overestimate the 
effect of diffusion on $R_\mathrm{AA}$ only when the system approaches the initialization time of hydro,
while in the very early stage the purely diffusive motion is a fair approximation.
Anyway, a more thorough study of the gluon 
radiation has to be pursued and this will be the subject of a forthcoming publication.

We suggest that the evolution of 
heavy quarks in the early stages of high energy nuclear collisions 
is important as it can alter the FFNLO initialization commonly used to study the heavy quarks
dynamics in these collisions. We remark that a simultaneous description of heavy quark
$R_{AA}$ and $v_2$ is a top challenge to almost all the models on heavy quark dynamics:
the inclusion of a pre-equilibrium phase might  improve the situation,
offering a better understanding of this puzzle~\cite{Sun:2019fud}.
This present work will further boost the phenomenology  as well as our understanding of the experimental results.

\begin{acknowledgements}
M. R. acknowledges Navid Abbasi, Gabriele Coci, Gianluca Giuliani, Feng Li, David Mueller,
John Petrucci, Yifeng Sun, Shen-Song Wan and Bo-Nan Zhang   for inspiration,
discussions and comments.
The work of M. R. and S. K. D. is supported by the National Science Foundation of China (Grants No. 11805087 and No. 11875153)
and by the Fundamental Research Funds for the Central Universities (grant number 862946).
The work of J. H. Liu is supported by China Scholarship Council (scholarship number 201806180032).
\end{acknowledgements}

\appendix

\section{Scaling to $SU(2)$ parameters}
In this section we present the result for $R_\mathrm{pPb}$ we obtain if we rescale the $g^2\mu$ of the MV model
from the $SU(3)$ value to the $SU(2)$ one: our goal is to show that this rescaling does not affect substantially the modification
factors computed in the article.

In order to rescale the $g^2 \mu$, 
we take the saturation momentum, $Q_s$, as independent on $N_c$, since $Q_s$ depends on the thickness function of a nucleus 
therefore it is sensitive to how nucleons are distributed, while the number of colors has to do with the substructure of the nucleons. 
On the other hand, the $g^2 \mu$ measures the density of the color charges at the quark-gluon level therefore it is sensitive to $N_c$. 
In the literature, a relation between the $Q_s$ and $g^2 \mu$ can be found in \cite{Lappi:2008eq}, 
which shows an explicit dependence on $N_c$ that neglecting logarithm is $Q_s^2/(g^2 \mu)^2 \propto N_c$. 
Leaving the $Q_s$ fixed, we change the ratio $g^2 \mu/Q_s$ by using the aforementioned relation:
 in particular, this scaling from 3 to 2 colors brings to $g^2 \mu_\mathrm{SU(2)} 
 \approx 1.22 \times g^2 \mu_\mathrm{SU(3)}$. 
Therefore, the $g^2 \mu$ has to be increased of approximately the 20$\%$.

\begin{figure}[t!]
\begin{center}
\includegraphics[width=0.45\textwidth]{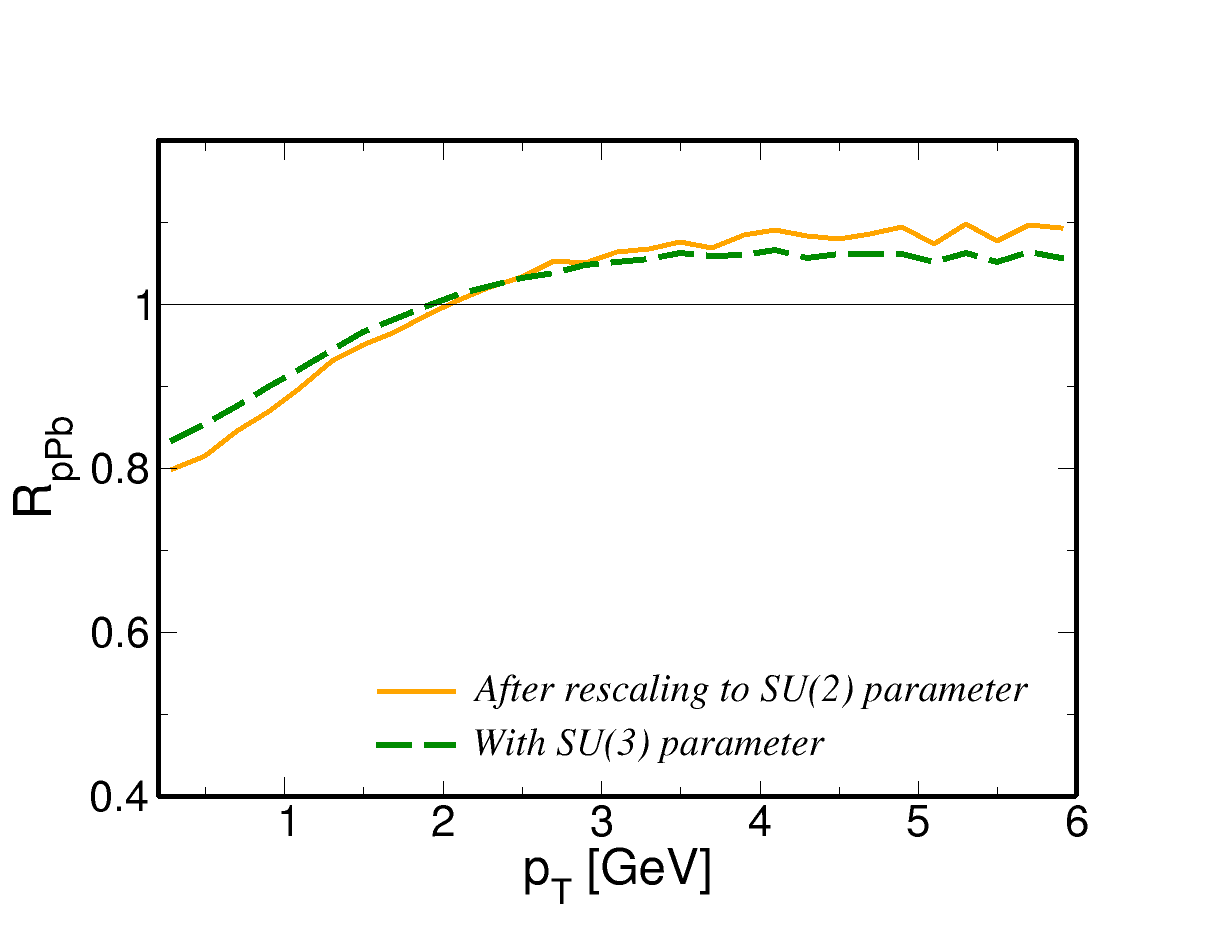}
\end{center}
\caption{\label{su2}
Comparison of the nuclear modification factor for the set up of p-Pb collisions,
obtained for the $g^2\mu$ of $SU(3)$ used in the main text, and for the $g^2\mu$ rescaled
to the $SU(2)$ case.}
\end{figure}

In Fig.~\ref{su2}  we compare the $R_\mathrm{pPb}$ 
obtained with the two values of the parameter: the $N_c=3$ case, which we use in the main text,
 and the $N_c=2$ obtained rescaling the $g^2\mu$ as outlined above. 
 We find that the net effect is less than 10$\%$. 
We have checked that this modest shift occurs also in the other cases studied in the article.


\begin{thebibliography}{99}

\bibitem{McLerran:1993ni}
  L.~D.~McLerran and R.~Venugopalan,
  Phys.\ Rev.\ D {\bf 49}, 2233 (1994)
  [hep-ph/9309289].
\bibitem{McLerran:1993ka}
  L.~D.~McLerran and R.~Venugopalan,
  Phys.\ Rev.\ D {\bf 49}, 3352 (1994)
  [hep-ph/9311205].
\bibitem{McLerran:1994vd}
  L.~D.~McLerran and R.~Venugopalan,
  Phys.\ Rev.\ D {\bf 50}, 2225 (1994)
  [hep-ph/9402335].

\bibitem{Gelis:2010nm}
  F.~Gelis, E.~Iancu, J.~Jalilian-Marian and R.~Venugopalan,
  Ann.\ Rev.\ Nucl.\ Part.\ Sci.\  {\bf 60}, 463 (2010).
\bibitem{Iancu:2003xm}
  E.~Iancu and R.~Venugopalan,
  In *Hwa, R.C. (ed.) et al.: Quark gluon plasma* 249-3363.
\bibitem{McLerran:2008es}
  L.~McLerran,
  arXiv:0812.4989 [hep-ph];
  hep-ph/0402137.

\bibitem{Gelis:2012ri}
  F.~Gelis,
  Int.\ J.\ Mod.\ Phys.\ A {\bf 28}, 1330001 (2013).

\bibitem{Kovner:1995ja}
  A.~Kovner, L.~D.~McLerran and H.~Weigert,
  Phys.\ Rev.\ D {\bf 52}, 6231 (1995)
  doi:10.1103/PhysRevD.52.6231
  [hep-ph/9502289].

\bibitem{Kovner:1995ts}
  A.~Kovner, L.~D.~McLerran and H.~Weigert,
  Phys.\ Rev.\ D {\bf 52}, 3809 (1995)
  doi:10.1103/PhysRevD.52.3809
  [hep-ph/9505320].

\bibitem{Gyulassy:1997vt}
  M.~Gyulassy and L.~D.~McLerran,
  Phys.\ Rev.\ C {\bf 56}, 2219 (1997)
  doi:10.1103/PhysRevC.56.2219
 [nucl-th/9704034].


\bibitem{Lappi:2006fp}
  T.~Lappi and L.~McLerran,
  Nucl.\ Phys.\ A {\bf 772}, 200 (2006)
  doi:10.1016/j.nuclphysa.2006.04.001
  [hep-ph/0602189].


\bibitem{Fries:2006pv}
  R.~J.~Fries, J.~I.~Kapusta and Y.~Li,
  nucl-th/0604054.

\bibitem{Chen:2015wia}
  G.~Chen, R.~J.~Fries, J.~I.~Kapusta and Y.~Li,
  Phys.\ Rev.\ C {\bf 92}, no. 6, 064912 (2015)
  doi:10.1103/PhysRevC.92.064912
  [arXiv:1507.03524 [nucl-th]].

\bibitem{Krasnitz:2000gz}
  A.~Krasnitz and R.~Venugopalan,
  Phys.\ Rev.\ Lett.\  {\bf 86}, 1717 (2001)
  doi:10.1103/PhysRevLett.86.1717
  [hep-ph/0007108].


\bibitem{Krasnitz:2001qu}
  A.~Krasnitz, Y.~Nara and R.~Venugopalan,
  Phys.\ Rev.\ Lett.\  {\bf 87}, 192302 (2001)
  doi:10.1103/PhysRevLett.87.192302
  [hep-ph/0108092].


\bibitem{Krasnitz:2003jw}
  A.~Krasnitz, Y.~Nara and R.~Venugopalan,
  Nucl.\ Phys.\ A {\bf 727}, 427 (2003)
  doi:10.1016/j.nuclphysa.2003.08.004
  [hep-ph/0305112].


\bibitem{Fukushima:2006ax}
  K.~Fukushima, F.~Gelis and L.~McLerran,
  Nucl.\ Phys.\ A {\bf 786}, 107 (2007)
  doi:10.1016/j.nuclphysa.2007.01.086
  [hep-ph/0610416].


\bibitem{Fujii:2008km}
  H.~Fujii, K.~Fukushima and Y.~Hidaka,
  Phys.\ Rev.\ C {\bf 79}, 024909 (2009)
  doi:10.1103/PhysRevC.79.024909
  [arXiv:0811.0437 [hep-ph]].


\bibitem{Fukushima:2013dma}
  K.~Fukushima,
  Phys.\ Rev.\ C {\bf 89}, no. 2, 024907 (2014)
  doi:10.1103/PhysRevC.89.024907
  [arXiv:1307.1046 [hep-ph]].



\bibitem{Romatschke:2005pm}
  P.~Romatschke and R.~Venugopalan,
  Phys.\ Rev.\ Lett.\  {\bf 96}, 062302 (2006)
  doi:10.1103/PhysRevLett.96.062302
  [hep-ph/0510121].


\bibitem{Romatschke:2006nk}
  P.~Romatschke and R.~Venugopalan,
  Phys.\ Rev.\ D {\bf 74}, 045011 (2006)
  doi:10.1103/PhysRevD.74.045011
  [hep-ph/0605045].


\bibitem{Fukushima:2011nq}
  K.~Fukushima and F.~Gelis,
  Nucl.\ Phys.\ A {\bf 874}, 108 (2012.
  doi:10.1016/j.nuclphysa.2011.11.003
  [arXiv:1106.1396 [hep-ph]].

\bibitem{Iida:2014wea}
  H.~Iida, T.~Kunihiro, A.~Ohnishi and T.~T.~Takahashi,
  arXiv:1410.7309 [hep-ph].


\bibitem{Gelis:2013rba}
  T.~Epelbaum and F.~Gelis,
  Phys.\ Rev.\ Lett.\  {\bf 111}, 232301 (2013)
  doi:10.1103/PhysRevLett.111.232301
  [arXiv:1307.2214 [hep-ph]].

\bibitem{Epelbaum:2013waa}
  T.~Epelbaum and F.~Gelis,
  Phys.\ Rev.\ D {\bf 88}, 085015 (2013)
  doi:10.1103/PhysRevD.88.085015
  [arXiv:1307.1765 [hep-ph]].

\bibitem{Ryblewski:2013eja}
  R.~Ryblewski and W.~Florkowski,
  Phys.\ Rev.\ D {\bf 88}, 034028 (2013)
  doi:10.1103/PhysRevD.88.034028
  [arXiv:1307.0356 [hep-ph]].

\bibitem{Ruggieri:2015yea}
  M.~Ruggieri, A.~Puglisi, L.~Oliva, S.~Plumari, F.~Scardina and V.~Greco,
  Phys.\ Rev.\ C {\bf 92}, 064904 (2015)
  doi:10.1103/PhysRevC.92.064904
  [arXiv:1505.08081 [hep-ph]].


\bibitem{Tanji:2011di}
  N.~Tanji and K.~Itakura,
  Phys.\ Lett.\ B {\bf 713}, 117 (2012)
  doi:10.1016/j.physletb.2012.05.043
  [arXiv:1111.6772 [hep-ph]].


\bibitem{Berges:2012cj}
  J.~Berges and S.~Schlichting,
  Phys.\ Rev.\ D {\bf 87}, no. 1, 014026 (2013)
  doi:10.1103/PhysRevD.87.014026
  [arXiv:1209.0817 [hep-ph]].


\bibitem{Berges:2013fga}
  J.~Berges, K.~Boguslavski, S.~Schlichting and R.~Venugopalan,
  Phys.\ Rev.\ D {\bf 89}, no. 11, 114007 (2014)
  doi:10.1103/PhysRevD.89.114007
  [arXiv:1311.3005 [hep-ph]].


\bibitem{Berges:2013lsa}
  J.~Berges, K.~Boguslavski, S.~Schlichting and R.~Venugopalan,
  JHEP {\bf 1405}, 054 (2014)
  doi:10.1007/JHEP05(2014)054
  [arXiv:1312.5216 [hep-ph]].

\bibitem{Berges:2013eia}
  J.~Berges, K.~Boguslavski, S.~Schlichting and R.~Venugopalan,
  Phys.\ Rev.\ D {\bf 89}, no. 7, 074011 (2014)
  doi:10.1103/PhysRevD.89.074011
  [arXiv:1303.5650 [hep-ph]].






\bibitem{Ruggieri:2017ioa}
  M.~Ruggieri, L.~Oliva, G.~X.~Peng and V.~Greco,
  Phys.\ Rev.\ D {\bf 97}, no. 7, 076004 (2018)
  doi:10.1103/PhysRevD.97.076004
  [arXiv:1707.07956 [nucl-th]].




\bibitem{Eskola:2009uj}
  K.~J.~Eskola, H.~Paukkunen and C.~A.~Salgado,
  JHEP {\bf 0904}, 065 (2009)
  doi:10.1088/1126-6708/2009/ 04/065
  [arXiv:0902.4154 [hep-ph]].


\bibitem{Rezaeian:2012ye}
  A.~H.~Rezaeian,
  Phys.\ Lett.\ B {\bf 718}, 1058 (2013)
  doi:10.1016/j.physletb.2012.11.066
  [arXiv:1210.2385 [hep-ph]].




\bibitem{Fujii:2013yja}
  H.~Fujii and K.~Watanabe,
  Nucl.\ Phys.\ A {\bf 920}, 78 (2013)
  doi:10.1016/j.nuclphysa.2013.10.006
  [arXiv:1308.1258 [hep-ph]].


\bibitem{Ducloue:2015gfa}
  B.~Ducloué, T.~Lappi and H.~Mäntysaari,
  Phys.\ Rev.\ D {\bf 91}, no. 11, 114005 (2015)
  doi:10.1103/PhysRevD.91.114005
  [arXiv:1503.02789 [hep-ph]].





\bibitem{Albacete:2013ei}
  J.~L.~Albacete {\it et al.},
  Int.\ J.\ Mod.\ Phys.\ E {\bf 22}, 1330007 (2013)
  doi:10.1142/S0218301313300075
  [arXiv:1301.3395 [hep-ph]].


\bibitem{Albacete:2016veq}
  J.~L.~Albacete {\it et al.},
  Int.\ J.\ Mod.\ Phys.\ E {\bf 25}, no. 9, 1630005 (2016)
  doi:10.1142/S0218301316300058
  [arXiv:1605.09479 [hep-ph]].
















\bibitem{Ruggieri:2019zos}
  M.~Ruggieri, M.~Frasca and S.~K.~Das,
  Chin.\ Phys.\ C {\bf 43}, no. 9, 094105 (2019)


\bibitem{Sun:2019fud} 
  Y.~Sun, G.~Coci, S.~K.~Das, S.~Plumari, M.~Ruggieri and V.~Greco,
  Phys.\ Lett.\ B {\bf 798}, 134933 (2019)
  doi:10.1016/j.physletb.2019.134933
  [arXiv:1902.06254 [nucl-th]].
















\bibitem{Prino:2016cni}
  F.~Prino and R.~Rapp,
  J.\ Phys.\ G {\bf 43}, no. 9, 093002 (2016)


\bibitem{Andronic:2015wma}
  A.~Andronic {\it et al.},
  Eur.\ Phys.\ J.\ C {\bf 76}, no. 3, 107 (2016)

\bibitem{Rapp:2018qla}
  R.~Rapp {\it et al.},
  Nucl.\ Phys.\ A {\bf 979}, 21 (2018)


\bibitem{Aarts:2016hap}
  G.~Aarts {\it et al.},
  Eur.\ Phys.\ J.\ A {\bf 53}, no. 5, 93 (2017)



\bibitem{Dong:2019unq}
  X.~Dong and V.~Greco,
  Prog.\ Part.\ Nucl.\ Phys.\  {\bf 104}, 97 (2019).


\bibitem{Cao:2018ews}
  S.~Cao {\it et al.},
  Phys.\ Rev.\ C {\bf 99}, no. 5, 054907 (2019)




\bibitem{Das:2016cwd}
  S.~K.~Das, S.~Plumari, S.~Chatterjee, J.~Alam, F.~Scardina and V.~Greco,
  Phys.\ Lett.\ B {\bf 768}, 260 (2017)




\bibitem{Das:2015ana}
  S.~K.~Das, F.~Scardina, S.~Plumari and V.~Greco,
  Phys.\ Lett.\ B {\bf 747}, 260 (2015)

\bibitem{Das:2017dsh}
  S.~K.~Das, M.~Ruggieri, F.~Scardina, S.~Plumari and V.~Greco,
  J.\ Phys.\ G {\bf 44}, no. 9, 095102 (2017)



\bibitem{Das:2015aga}
  S.~K.~Das, M.~Ruggieri, S.~Mazumder, V.~Greco and J.~e.~Alam,
  J.\ Phys.\ G {\bf 42}, no. 9, 095108 (2015)


\bibitem{Beraudo:2015wsd}
  A.~Beraudo, A.~De Pace, M.~Monteno, M.~Nardi and F.~Prino,
  JHEP {\bf 1603}, 123 (2016)
  doi:10.1007/JHEP03(2016)123
  [arXiv:1512.05186 [hep-ph]].

\bibitem{Xu:2015iha}
  Y.~Xu, S.~Cao, G.~Y.~Qin, W.~Ke, M.~Nahrgang, J.~Auvinen and S.~A.~Bass,
  Nucl.\ Part.\ Phys.\ Proc.\  {\bf 276-278}, 225 (2016)
  doi:10.1016/j.nuclphysbps.2016.05.050
  [arXiv:1510.07520 [nucl-th]].

\bibitem{Ozvenchuk:2017ojj}
  V.~Ozvenchuk, J.~Aichelin, P.~B.~Gossiaux, B.~Guiot, M.~Nahrgang and K.~Werner,
  J.\ Phys.\ Conf.\ Ser.\  {\bf 779}, no. 1, 012033 (2017).
  doi:10.1088/1742-6596/779/1/012033


\bibitem{Das:2013kea}
  S.~K.~Das, F.~Scardina, S.~Plumari and V.~Greco,
  Phys.\ Rev.\ C {\bf 90}, 044901 (2014)
  doi:10.1103/PhysRevC.90.044901
  [arXiv:1312.6857 [nucl-th]].


\bibitem{Chandra:2015gma}
  V.~Chandra and S.~K.~Das,
  Phys.\ Rev.\ D {\bf 93}, no. 9, 094036 (2016)
  doi:10.1103/PhysRevD.93.094036
  [arXiv:1506.07805 [nucl-th]].

\bibitem{Mrowczynski:2017kso}
  S.~Mrowczynski,
  Eur.\ Phys.\ J.\ A {\bf 54}, no. 3, 43 (2018)
  doi:10.1140/epja/i2018-12478-5
  [arXiv:1706.03127 [nucl-th]].


\bibitem{Ruggieri:2018rzi} 
  M.~Ruggieri and S.~K.~Das,
  Phys.\ Rev.\ D {\bf 98}, no. 9, 094024 (2018)
  doi:10.1103/PhysRevD.98.094024
  [arXiv:1805.09617 [nucl-th]].

\bibitem{Ruggieri:2018ies} 
  M.~Ruggieri and S.~K.~Das,
  EPJ Web Conf.\  {\bf 192}, 00017 (2018)
  doi:10.1051/epjconf/201819200017
  [arXiv:1809.07915 [nucl-th]].


\bibitem{Abelev:2014hha}
  B.~B.~Abelev {\it et al.} [ALICE Collaboration],
  Phys.\ Rev.\ Lett.\  {\bf 113}, no. 23, 232301 (2014)
  doi:10.1103/PhysRevLett.113.232301
  [arXiv:1405.3452 [nucl-ex]].


\bibitem{Aaij:2017gcy}
  R.~Aaij {\it et al.} [LHCb Collaboration],
  JHEP {\bf 1710}, 090 (2017)
  doi:10.1007/JHEP10(2017)090
  [arXiv:1707.02750 [hep-ex]].



\bibitem{Kovchegov:1996ty}
  Y.~V.~Kovchegov,
  Phys.\ Rev.\ D {\bf 54}, 5463 (1996)
  doi:10.1103/PhysRevD.54.5463
  [hep-ph/9605446].


\bibitem{Lappi:2007ku}
  T.~Lappi,
  Eur.\ Phys.\ J.\ C {\bf 55}, 285 (2008)
  doi:10.1140/epjc/s10052-008-0588-4
  [arXiv:0711.3039 [hep-ph]].



\bibitem{Schenke:2014zha}
  B.~Schenke and R.~Venugopalan,
  Phys.\ Rev.\ Lett.\  {\bf 113}, 102301 (2014)
  doi:10.1103/PhysRevLett.113.102301
  [arXiv:1405.3605 [nucl-th]].

\bibitem{Schenke:2015aqa}
  B.~Schenke, S.~Schlichting and R.~Venugopalan,
  Phys.\ Lett.\ B {\bf 747}, 76 (2015)
  doi:10.1016/j.physletb.2015.05.051
  [arXiv:1502.01331 [hep-ph]].

\bibitem{Mantysaari:2017cni}
  H.~Mäntysaari, B.~Schenke, C.~Shen and P.~Tribedy,
  Phys.\ Lett.\ B {\bf 772}, 681 (2017)
  doi:10.1016/j.physletb.2017.07.038
  [arXiv:1705.03177 [nucl-th]].


\bibitem{Mantysaari:2016jaz}
  H.~Mäntysaari and B.~Schenke,
  Phys.\ Rev.\ D {\bf 94} (2016) no.3,  034042
  doi:10.1103/PhysRevD.94.034042
  [arXiv:1607.01711 [hep-ph]].



\bibitem{GolecBiernat:1999qd}
  K.~J.~Golec-Biernat and M.~Wusthoff,
  Phys.\ Rev.\ D {\bf 60}, 114023 (1999)
  doi:10.1103/PhysRevD.60.114023
  [hep-ph/9903358].


\bibitem{GolecBiernat:1998js}
  K.~J.~Golec-Biernat and M.~Wusthoff,
  Phys.\ Rev.\ D {\bf 59}, 014017 (1998)
  doi:10.1103/PhysRevD.59.014017
  [hep-ph/9807513].


\bibitem{Kovchegov:2012mbw}
  Y.~V.~Kovchegov and E.~Levin,
  {\it Quantum chromodynamics at high energy,}
  Camb.\ Monogr.\ Part.\ Phys.\ Nucl.\ Phys.\ Cosmol.\  {\bf 33} (2012).

\bibitem{Kowalski:2007rw}
  H.~Kowalski, T.~Lappi and R.~Venugopalan,
  Phys.\ Rev.\ Lett.\  {\bf 100}, 022303 (2008)
  doi:10.1103/PhysRevLett.100.022303
  [arXiv:0705.3047 [hep-ph]].

\bibitem{Armesto:2004ud}
  N.~Armesto, C.~A.~Salgado and U.~A.~Wiedemann,
  Phys.\ Rev.\ Lett.\  {\bf 94}, 022002 (2005)
  doi:10.1103/PhysRevLett.94.022002
  [hep-ph/0407018].

\bibitem{Freund:2002ux}
  A.~Freund, K.~Rummukainen, H.~Weigert and A.~Schafer,
  Phys.\ Rev.\ Lett.\  {\bf 90}, 222002 (2003)
  doi:10.1103/PhysRevLett.90.222002
  [hep-ph/0210139].




\bibitem{Albacete:2012xq}
  J.~L.~Albacete, A.~Dumitru, H.~Fujii and Y.~Nara,
  Nucl.\ Phys.\ A {\bf 897}, 1 (2013)
  doi:10.1016/j.nuclphysa.2012.09.012
  [arXiv:1209.2001 [hep-ph]].


\bibitem{FONLL}
  M.~Cacciari, M.~Greco and P.~Nason,
  JHEP {\bf 9805} (1998) 007 [arXiv:hep-ph/9803400];
  M.~Cacciari, S.~Frixione and P.~Nason,
  JHEP {\bf 0103} (2001) 006 [arXiv:hep-ph/0102134].


\bibitem{Cacciari:2012ny}
  M.~Cacciari, S.~Frixione, N.~Houdeau, M.~L.~Mangano, P.~Nason and G.~Ridolfi,
  JHEP {\bf 1210} (2012) 137 [arXiv:1205.6344 [hep-ph]].


\bibitem{Cacciari:2015fta}
  M.~Cacciari, M.~L.~Mangano and P.~Nason,
  arXiv:1507.06197 [hep-ph].


\bibitem{Wong:1970fu}
  S.~K.~Wong,
  Nuovo Cim.\ A {\bf 65}, 689 (1970).
  doi:10.1007/BF02892134


\bibitem{Boozer}
A. D. Boozer, Am. J. Phys. 79 (9), September 2011.

\bibitem{Pet} C. Peterson {\it et al.},  Phys. Rev. D {\bf 27}, 105 (1983).




\bibitem{Scardina:2017ipo}
  F.~Scardina, S.~K.~Das, V.~Minissale, S.~Plumari and V.~Greco,
  Phys.\ Rev.\ C {\bf 96}, no. 4, 044905 (2017)
  doi:10.1103/PhysRevC.96.044905
  [arXiv:1707.05452 [nucl-th]].

\bibitem{Plumari:2017ntm}
  S.~Plumari, V.~Minissale, S.~K.~Das, G.~Coci and V.~Greco,
  Eur.\ Phys.\ J.\ C {\bf 78}, no. 4, 348 (2018)
  doi:10.1140/epjc/s10052-018- 5828-7
  [arXiv:1712.00730 [hep-ph]].



\bibitem{Lappi:2008eq}
T.~Lappi,
J. Phys. G \textbf{35}, 104052 (2008)
doi:10.1088/0954-3899/35/10/104052
[arXiv:0804.2338 [hep-ph]].

 





\end{thebibliography}
\end{document}